\documentclass[preprint,showpacs,preprintnumbers,superscriptaddress,amsmath,amssymb,floatfix,prl]{revtex4}

\usepackage{graphicx}
\usepackage{color}
\usepackage{amsmath}
\usepackage{color}
\usepackage{dcolumn}
\usepackage{bm}

\def\Wcm2{\mbox{W cm}^{-2}\mu m ^2 }
 \def\fs{\mbox{fs}}

\begin{document}

\title{ Simple Scalings for Various Regimes of Electron Acceleration in Surface Plasma Waves}
\author{C.~Riconda}
\affiliation{LULI, Sorbonne Universit\'e, Universit\'e Pierre et Marie Curie, Ecole Polytechnique, CNRS UMR 7605, CEA, Paris,  75005 France}
\author{M.~Raynaud}
\affiliation{Laboratoire des Solides Irradi\'es, CNRS UMR 7642, CEA-DSM-IRAMIS, Ecole Polytechnique, Universit\'e Paris-Saclay, 91128 Palaiseau, France}
\author{T.~Vialis}
\affiliation{LULI, Sorbonne Universit\'e, Universit\'e Pierre et Marie Curie, Ecole Polytechnique, CNRS UMR 7605, CEA, Paris,  75005 France}
\author{M.~Grech}
\affiliation{LULI, CNRS UMR 7605, Universit\'e Pierre et Marie Curie, Ecole Polytechnique, CEA, 91128 Palaiseau, France}

\date{\today}

\begin{abstract}
Different electron acceleration regimes in the evanescent field of a surface plasma wave are studied by considering the interaction of a test electron with the 
high-frequency electromagnetic field of a surface wave. The non-relativistic and relativistic limits are investigated. Simple scalings are found demonstrating the possibility  to achieve  an efficient conversion of the surface wave field energy into electron kinetic energy.
This mechanism of electron acceleration can provide a high-frequency pulsed source of  relativistic electrons with a well defined energy. In  the relativistic limit, the most energetic electrons  are obtained in the so-called electromagnetic regime for surface waves. In this regime the particles are  accelerated to velocities larger than the wave phase velocity, mainly in the direction parallel to the plasma-vacuum interface.

\end{abstract}

\pacs{52.65.Cc,73.20.Mf,52.35.Mw}

\maketitle

\section{Introduction}

Electron acceleration by laser-plasma interaction has been studied extensively within the context of laser absorption by a plasma. It has also been studied for the development of techniques aimed at producing hot electrons in order to obtain improved energetic electron sources and to enhance secondary processes including X-ray, $\gamma$-ray, positron production and ion acceleration \cite{corde}. In under-dense plasmas, various methods of wakefield acceleration \cite{tajima:79,esarey:96,naseri:12} have been proposed and electrons up to $\sim 4.2~GeV$ have been experimentally observed in the optimal configuration \cite{leemans}. However, the electron acceleration mechanism invoked  involves high laser intensity, well into the relativistic regime, and short pulse duration $\tau_{L} = 40~fs$ resulting in a small current (total current of $\sim10 pC $). 
It is thus of interest to search for alternative configurations wherein the current can be enhanced.

In dense plasmas, an energetic electron population may be created by resonant absorption \cite{resabs:69}. In this case, the electrons are accelerated by the resonant plasma waves excited at densities around the critical density by the interaction of a ``long'' laser pulse with a gentle-gradient dense plasma.
With the development of lasers of high intensity ($I\lambda_0^2>10^{16}\Wcm2$), short pulse duration ($<100\fs$), and high contrast ($\sim 10^{12}$), new electron heating mechanisms have also been proposed \cite{brunel:87,kruer:85,skineffect1,skineffect2, jin}. They involve very sharp density gradients and over-dense plasmas which do not necessarily require a resonant plasma response. Among these mechanisms are vacuum heating \cite{brunel:87}, and the so-called ponderomotive or $\vec J \times \vec B$ heating \cite{kruer:85}. Numerical simulations \cite{gibbon:92} have shown that, in many cases, vacuum heating can be more efficient than resonant absorption. 
The advantage of these mechanisms is that they involve dense plasmas such that they bear the potential to generate very high currents.

In order to improve the electron acceleration to relativistic values by a laser interacting with an over-dense plasma, particular targets designs can be used \cite{kluge,jiang}. Also surface electron plasma waves resonantly excited by the laser on a structured target can be  used to reach that goal \cite{ESW2,bigon:13}. The electrons interact in this case with the surface plasma wave field that is highly localized at the vacuum-plasma interface and oscillates with the laser frequency. The local field amplitude is  higher than that of the incoming laser  such that the typical electron quiver motion is much faster than the electron thermal velocity. Experimental evidence for the feasibility to accelerate  electrons and ions  by this kind of scheme have recently been reported \cite{ceccotti:13}. 

Surface plasma waves can also be excited at the surface of a laser-produced ion channel (by a mechanism similar to wakefield acceleration). Their use has been proven to be an efficient means to improve fast electron generation \cite{naseri:12,naseri:13,willingale:11,willingale:13}.
Moreover, full 2D simulations of $\vec J \times \vec B$ heating \cite{macchi:01} have shown that these oscillations can be unstable and decay into a standing surface plasma wave, revealing at the same time the existence of a  non-linear mechanism for generation of surface plasma waves.

However, depending on the characteristics of the excited surface plasma waves the electron population can have very different features. In this paper we explore with a simple model the different acceleration regimes of an electron in the evanescent electromagnetic field of a surface plasma wave by considering the interaction of a test electron with the high-frequency surface wave fields. The non-relativistic and relativistic limits are investigated by means of $1D$ and $2D$ test particle simulations in order to find the optimal regime for an efficient conversion of the surface wave field energy into electron kinetic energy. The 1D dynamics calculations are mainly performed in order to analyze the motion of the electron in the direction perpendicular to the surface and its excursion into the evanescent surface plasma field. Some analysis of the $1D$ motion in the direction parallel to the plasma-vacuum surface is also performed.

After a brief summary of the structure of the surface plasma waves (section II), 
the electron acceleration mechanism proposed  is studied, first of all for non-relativistic 
over-dense plasmas (section III). The treatment of this case is based on classical concepts such as the idea of a ponderomotive potential \cite{Brucksbaum} and the separation of high-frequency and low-frequency effects, which allow an analytical 1D treatment (subsection III-A). This part is then complemented by full 2D test particle simulations in  subsection III-B. 
It permits to identify two different situations for the electron acceleration: the electromagnetic regime and the electrostatic regime. This non relativistic part of the study is also a tool to understand the results in the relativistic regime.
The next section (section IV) studies the relativistic limit where the quiver velocity becomes close to the velocity of light. First a numerical $1D$ study is  presented (subsection IV-A), as within this domain the possibilities for an analytical treatment  are limited.  This allows the determination of the different regimes, and of the optimal conditions in terms of electron acceleration. To complete the study, a full $2D$ numerical study is presented in subsection  IV-B. 
We find that the electromagnetic regime is the most efficient and taht an electron initially at rest can be self-injected and phase-locks on the vacuum plasma side through $\vec v \times \vec B$ mechanisms. This way the electron is  accelerated to velocities larger than the wave phase velocity, mainly in the direction parallel to the plasma-vacuum interface. The resulting electron total energy scales as $\gamma_\varphi a_{sw} mc^2$ with $a_{sw}=eE_{sw}/mc\omega$ where $E_{sw}$ is the SPW field component perpendicular to the surface of the plasma, $\gamma_{\varphi}=1/\sqrt{1 - (v_\varphi/c)^2}$ with $v_\varphi$ the phase velocity of the SPW and $\omega$ is the surface plasma wave frequency.

\section{Structure of the Surface Plasma waves}
 
We consider hereafter a $2D$ homogeneous plasma in the $(x,y)$ plane, which supports a surface plasma wave that propagates at the plasma-vacuum interface (along the $y$-direction). The vacuum is along $x<0$.  On  the vacuum side ($x<0$), the resonant surface wave field has  the form:
\begin{equation}\label{champvide}
   \begin{cases}
       E_{x,v}= -E_{sw} f(t,x) sin(\omega t- k_yy+\phi),\\
       E_{y,v}= E_{sw}\frac{1}{k_y L_{E,v}} f(t,x) cos(\omega t-k_yy+\phi),\\
       B_{z,v}= E_{sw}\frac{v_{\varphi}}{c} f(t,x) sin(\omega t-k_yy+\phi),\\
   \end{cases} 
\end{equation}      
with: 
$$f(t,x)=\exp(x/L_{E,v})\exp(-2 t^2/ \tau_{sw}^2),
 $$\\
\noindent while on the plasma side ($x>0$):
\begin{equation}\label{champplasma}
   \begin{cases}
       E_{x,p}= E_{sw} \frac{L_{E,p}}{L_{E,v}}g(t,x) sin(\omega t-  k_yy+\phi),\\
       E_{y,p}=E_{sw}\frac{1}{k_y L_{E,v}} g(t,x) cos(\omega t-k_yy+\phi),\\
       B_{z,p}=E_{sw}\frac{v_{\varphi}}{c} g(t,x) sin(\omega t- k_yy+\phi),\\
    \end{cases}
\end{equation}    
with:   
$$g(t,x)=\exp(-x/L_{E,p})\exp(-2 t^2/ \tau_{sw}^2),
      $$\\
\noindent where $E_{sw}$ is the maximum value of the electric field on the vacuum side  in the $x$-direction. In   the following this field will serve as  reference field. $\tau_{sw}$ is the mode lifetime, $\phi$ is the phase, $v_{\varphi}= \omega/k_y$ is the phase velocity of the wave and $L_{E,v}$ and $L_{E,p}$ are the evanescence length in the vacuum and plasma respectively. 
The expression for the  evanescence length in vacuum is given by 
$L_{E,v}= \sqrt{(k^2_y- \omega^2/c^2)^{-1}}$. The evanescence length within the plasma is typically significantly smaller than the evanescence length in vacuum \cite{kaw:70}. It is given by $L_{E,p}= \sqrt{(k^2_y- (\omega^2-\omega_{pe}^2)/c^2)^{-1}} < L_{E,v}$.
$\tau_{sw}$ will depend on the damping mechanisms that affects the plasma wave, such as  linear and non-linear Landau damping, and wave breaking effects. 
 
In the cold-plasma limit, the thermal corrections to the dispersion relation for the surface plasma 
waves can be neglected, and  in the non-relativistic limit we have \cite{kaw:70} the relation:
\begin{equation}\label{disp}
{k^2_y c^2 \over \omega^2} = {1-(\omega/\omega_{pe})^2
\over 1-2(\omega/\omega_{pe})^2 }, 
\end{equation} 
  
\noindent where $\omega^2_{pe}=4\pi n e^2/m$ is the electron plasma frequency, $n$ the electron density and $m$ the electron mass. By solving Eq. (\ref{disp}) for $\omega (k)$ it is easily seen that there is an upper limit for $\omega \le \omega_{pe}/\sqrt{2}$. The surface plasma waves which satisfy this dispersion relation are localized at the plasma surface.  

Let us now consider in the following some limiting situations for the surface plasma wave. The relevant parameter  to identify the different regimes for the electron plasma wave is $\omega/\omega_{pe}$. We will therefore consider different values of this parameter, and the corresponding wave vector that satisfies the dispersion relation Eq.(\ref{disp}).

In the limit $\omega\ll\omega_{pe}$, which is the so-called electromagnetic limit for the surface plasma waves, the magnetic field is of the same order of magnitude as the perpendicular electric field ($E_x$), while  the component of the electric field parallel to the surface ($E_y$) is negligible (see Eq.\ref{champvide}, which in this limit yields $|E_y| \sim \frac{\omega}{\omega_{pe}} |E_x|$). Thus the electric field is mainly in the $x$ direction (i.e. perpendicular to the plasma-vacuum interface). It should be noted that  in this case two different situations can be expected, depending on the field intensity. At low field intensity, it can be anticipated that the role of the magnetic field will be negligible,  such that a 1D model can be used, wherein the motion only takes place along $x$. In the opposite case that the value of the fields is large enough to accelerate the electrons into the relativistic regime, the magnetic field contribution will play an important role. In the electromagnetic limit where $\omega\ll\omega_{pe}$, the wave phase velocity along the surface, $v_\varphi=\omega/k_y$, is  less than but of the same order of magnitude as the velocity of light. Moreover the evanescence length of the wave in vacuum can be quite large: the smaller the value $\omega/\omega_{pe}$, the larger the vacuum evanescence length. For example,  if  $\omega/\omega_{pe} = 0.05$ we have from Eq. (\ref{disp}) $K \equiv k_y^2 c^2/\omega^2 = 1.0025$, and $L_{E,v} = \lambda_0/2\pi\sqrt{K-1} = 3.18 \lambda_0$, for $\lambda_0=2 \pi c/\omega $. If instead we consider  $\omega/\omega_{pe} = 0.22$, we have $K = 1.053$, and $L_{E,v} = 0.7 \lambda_0$.  
As we will see in the following, large phase velocities $v_\varphi=c/\sqrt{K}$, and large evanescent lengths $L_{E,v}$ are favorable situations for efficient electron acceleration within the highly relativistic regime.  Note that in the electromagnetic limit the evanescence length in the plasma is much less than the evanescence length in vacuum. In particular in the limit $\omega\ll\omega_{pe}$ we have $L_{E,p} \sim c/\omega_{pe}$. 

On the other hand, in the limit $\omega\sim\omega_{pe}/\sqrt{2}$, we are close to the electrostatic limit where the magnetic field is always negligible. The two components of the electric field are of the same order of magnitude and the evanescent length is very small. Typically we always have $L_{E,v},L_{E,p} \ll \lambda_0$. 
In particular, if $\omega= 0.7 ~\omega_{pe} \sim \omega_{pe}/\sqrt{2}$, we have $K = 25.5$, and $L_{E,v}=0.032 \lambda_0$. In this case, the wave phase velocity along the surface is much smaller than the velocity of light.

We have thus identified two important parameters characterizing the surface plasma waves, the phase velocity, $v_\varphi$, along the surface and the evanescent length $L_{E,v}$ perpendicular to the surface. 
These two parameters will allow us to define different regimes governing the particle acceleration.
This will be done in the next section, where we will concentrate on 
the solution of the wave on the vacuum side, since it corresponds to a more favorable situation for electron acceleration. 
In table I, we have reported these parameters for different values of $\omega/\omega_{pe}$ to be discussed below. 

\begin{table}
\caption{\label{runs} Summary of the parameters for the surface plasma wave with the following definitions: 
$K=(k_y c/\omega)^2$, $L_{E,v}= \sqrt{(k^2_y- \omega^2/c^2)^{-1}}$, $v_\varphi=c/\sqrt{K}$ and $\gamma_{\varphi}=1/\sqrt{1 - (v_\varphi/c)^2}$. The different values of $\omega/\omega_{pe}$ correspond to different 
regimes for the surface waves.
\\
  }
\begin{tabular}{|c|c|c|c|c|c|}
\hline
$\omega/\omega_{pe}$& $K$ & ${L_{E,v}}/{\lambda_0}$ & ${E_{y,v}}/{E_{x,v}}$ &$v_\varphi/c$ & $\gamma_{\varphi}$ \\
\hline
0.7 & 25.5  & 0.032  &  0.98  & 0.198 & 1.02 \\
0.6 & 2.28  & 0.14  &  0.75 &  0.662 &1.33 \\ 
0.22 & 1.053 & 0.69  & 0.2  & 0.974 &4.54 \\
0.05 & 1.0025 & 3.18  &  0.05  & 0.998 & 20 \\
\hline
\end{tabular} 
\end{table}

\section{Non-relativistic Limit}

We first analyse the case of low surface plasma field intensity such that $a_{sw}=eE_{sw}/mc\omega \ll 1$. Relativistic effects are then negligible. Let us recall here that $E_{sw}$ is the maximum value of the electric SPW field component perpendicular to the surface in vacuum, $E_{x,v}$, and that the perpendicular component of the field is significantly reduced inside the plasma. For this reason we focus on the motion of the electrons in the vacuum side. As we are typically in the limit $\tau_{sw} \gg 2\pi / \omega$ we will have a slow time evolution combined with a fast time variation due to the high-frequency collective electron oscillations.

\subsection{Acceleration perpendicular to the surface}
We start the study of the motion of the electrons by considering only the behavior in the perpendicular direction, $x$, with the aim  of highlighting the role of the evanescence length of the field. In this case, we consider a fixed value of $y$, viz. $y=0$, and we neglect the contributions of $E_{y,v}$ and $B_{z,v}$. This is  quite good an approximation
 for the electromagnetic case since $E_{y,v}/E_{x,v} \ll 1$ (see table I) and the electron velocity is much smaller than the velocity of light.
Associated to the maximum field amplitude we can define the electron quiver velocity $v_{osc}= eE_{sw}/\gamma_{osc}m\omega$, where $\gamma_{osc}=\sqrt{1 + (eE_{sw}/c m\omega)^2}$. In the non-relativistic limit considered here $\gamma_{osc} \sim 1$, but we introduce it here  in order to formulate the most general definition of $v_{osc}$.

The high-frequency motion of a single electron in the electric field will be 
characterized by the length $\Lambda= v_{osc}/\omega$. This quantity should be compared 
with the length scale at which the field $L_{E,v}$ varies and we define hereafter the ratio $R_L=L_{E,v}/\Lambda$. 

If $R_L \gg 1$, that is when $v_{osc}$ is not too large, the electrons will have time to perform many oscillations before leaving the resonant field. This  is the relevant physical regime at low laser intensities where relativistic effects are negligible. In this case the exact values of $L_{E,v}$, or $\omega/\omega_{pe}$ do not matter, since 
to a good approximation
all frequencies will correspond  to the same physical situation, as long as $R_L \gg 1$. 
In this limit, the electrons undergo the effect of the ponderomotive force, which implies an averaging over the fast oscillatory motion in the field gradient. Hence, the typical kinetic energy acquired by an electron can be equal to the ponderomotive potential $U_{osc} = (1/4) m_e v_{osc}^2$. 
 
In fact, the lifetime $\tau_{sw}$ of the surface plasma wave may play an important role in determining the amount of net kinetic energy that can be gained by the electrons within the surface plasma wave. It has been shown in particular\cite{kuper:01}, that only a partial conversion of the potential energy into kinetic energy can occur
 if $\tau_{sw}$ is shorter than the time ($\sim L_{E,v}/v_{osc}$) needed by the electron to explore the whole spatial extension of the surface plasma wave field. In this case, the net kinetic energy gained by the electron  will be only a fraction of $U_{osc}$. 

There exists an additional source for electron acceleration, that  is not of ponderomotive origin 
inasmuch it does not specifically involve  a low-frequency time scale or a spatial 
dependence of the field.  At the moment they enter into the electric field of the surface wave\cite{kuper:01}, the electrons can also gain some kinetic energy on a sub-period time scale, that is, before even feeling the effect of the ponderomotive force. This extra kinetic energy may be simply related to the motion of the electron entering in a spatially constant oscillatory field and is, therefore, a strong function of the entry phase of the electron in the surface wave. The variations in the entry phase reflect  the variations of the entry times of the electron into the surface wave field. However in the following we keep the initial time constant equal to zero and we consider explicitly different values of the phase $\phi$.

A first value for the total kinetic energy acquired by an electron can be obtained from the zero order term in a series expansion for the  solution of the equations in the presence of an electric field gradient. Thus we can solve  the 
equations of motion of an electron with initial velocity $v=-v_0 $ at $t=0$ in a spatially constant field, whereby $v_0 $ is
of the order of the thermal velocity. After averaging over the fast motion, the resulting velocity as a function of the phase is given by:
\begin{equation}\label{vave}
<v(\phi)>_{h.f.}=-v_0+v_{osc}\cos\phi.
\end{equation}

\noindent Formally this is equivalent to having as initial velocity $<v(\varphi)>_{h.f.}$ instead of $-v_0$. The first order solution in the electric-field gradient, yields the ponderomotive force: At this stage
any other dependence on the phase disappears in the averaging procedure that is necessary to define the ponderomotive energy, such that the  final kinetic energy ($W_{K,f}$) of the electron is obtained by adding the ponderomotive potential to the kinetic energy associated with the velocity given above. By assuming that  the ponderomotive potential energy is completely converted into kinetic energy, we obtain:
\begin{equation}\label{kinen}
W_{K,f}= U_{osc} +2 U_{osc} \left(-{v_0 \over v_{osc}} + \cos\phi \right)^{2}.
\end{equation}
It is convenient to express Eq. (\ref{kinen}) in terms of the final value $v_f$ of the velocity acquired by the electrons: 
\begin{equation}\label{kabal}
{v_f \over v_{osc}} = \left(0.5+ 
\left(-{v_0 \over v_{osc}} + \cos\phi \right)^2\right)^{1/2}.
\end{equation}
This equation shows that the phase for which the electron obtains the maximum energy (best phase), is 
given by $\phi = \pi$. For this phase, the electron leaves the surface plasma wave field with a maximum kinetic energy such that $v_f$ is  greater than $ v_{osc}/\sqrt{2}$. Hence, 
due to this phase dependence, there will exist electrons which are accelerated to energies higher than the ponderomotive potential energy. Even if the contribution to the electron energy in this case
 is not uniquely of ponderomotive origin as explained above, we call this the ponderomotive regime.

In the following, we have  numerically solved the 1D equation of motion:
\begin{equation}
{d\overrightarrow{p} \over dt} = -e(\overrightarrow{E} + {\overrightarrow{v}\over c}\wedge\overrightarrow{B}) 
\end{equation} 
with $\overrightarrow{p}=\gamma m\overrightarrow{v}$
for an electron subjected to an external field ${\overrightarrow{E}}$, 
representing the field of a surface plasma wave, as defined in section I. The initial conditions are such that at $t=0$, $x=0, v=-v_0$, for $v_0$ corresponding to $100eV$, that corresponds to $v_0/v_{osc}= 0.16$.  The surface plasma wave lifetime is taken as $30\tau_0$ where $\tau_0=2\pi/\omega$. The electron motion is solved by using the Vay pusher \cite{vay}. We have adopted a value 
$E_{sw}=2.76\times 10^9 V/cm$ ($a_{sw}\sim 0.086$), for which $\gamma_{osc} \sim 1$. The numerical results are reported in fig.1 for $\omega/\omega_{pe} = 0.22$. For this case we have $R_L \sim 50$ and $L_{E,v} = 0.7 \mu m$. For comparison, the analytical curve obtained from Eq. (\ref{kabal}) is also reproduced  in the figure.

\begin{figure}
\includegraphics[height=5.5cm,width=6.5cm]{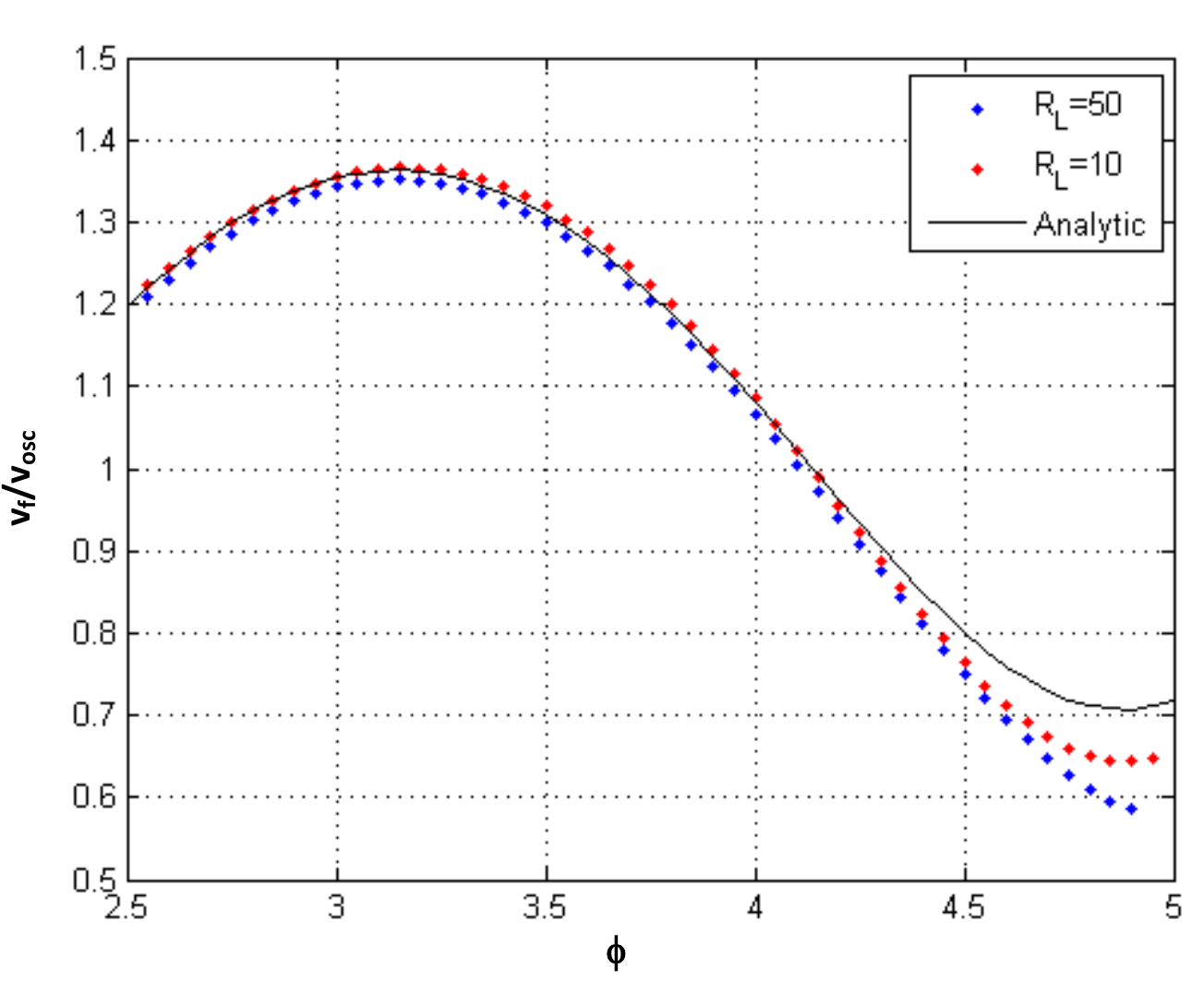}
\caption{
\label{fig:01} Non-relativistic limit: final  electron velocity $v_f/v_{osc}$ as a function of the entry phase $\phi$ in the SPW field. Here $\frac{1}{2}mv_0^2=100eV$ and $v_{osc}/c \sim a_{sw}= 0.086$. We plot equation (6) (black full line), and the numerical values for $R_L \sim 50$ ($\omega = 0.22 \omega_{pe}$) ({$\color{blue}\ast$}), and $R_L \sim 10$ ($\omega = 0.6 \omega_{pe}$)({$\color{red}\ast$}). Only the electrons in the vacuum side have been considered.}
\end{figure}

For the phase $\phi \sim \pi$, the numerical and theoretical results are in very good agreement. They  correspond to electrons that have acquired the possibility  to fully explore the field gradient
during their motion, after having been accelerated to their maximum energy. When the initial phase approaches $3\pi/2$, the discrepancy of the numerical curve with the theoretical prediction of Eq. (\ref{kabal}) grows larger. 
This prediction is obtained by assuming that the ponderomotive potential is completely converted into kinetic energy.  This can be explained by noticing that for the less favorable phase the electrons are 
moving slower such that they need more time to cross the field. Under these conditions, the fact that the wave has a finite lifetime  can no longer be neglected ($\omega\tau_{sw} \sim 180$) such that only a partial conversion of potential energy into kinetic energy can take place. This is verified by considering an intermediate case, with a shorter evanescence length $L_{E,v}$: for example $R_L \sim 10$ ($\omega= 0.6~\omega_{pe}$, $L_{E,v}=0.14 ~\mu m$). As expected, the  acceleration obtained and reported in fig.\ref{fig:01} is now closer to the theoretical prediction for $\phi \sim 3\pi/2$. 
In conclusion, we can say that the phase of the field experienced by the electron plays an essential part in determining the range of energy that can be acquired on traveling through the field. Moreover the role of the surface plasma wave lifetime is determined by the parameter  $L_{E,v}/v_{osc} \tau_{sw}$, and is negligible if this parameter is small.

\subsection{2D case}

We now consider the full $2D$ motion of the electron in the non-relativistic limit  in more detail. Note that in this case, the entry phase $\phi$ is both representative of electrons entering  the wave at different times or in different space locations in the $y$-direction. (In the simulation, only the electrons in the vacuum side have considered and we have taken $x=0$ and $y=0$ at $t=0$).

\begin{figure}
\includegraphics[height=5.5cm,width=7.cm]{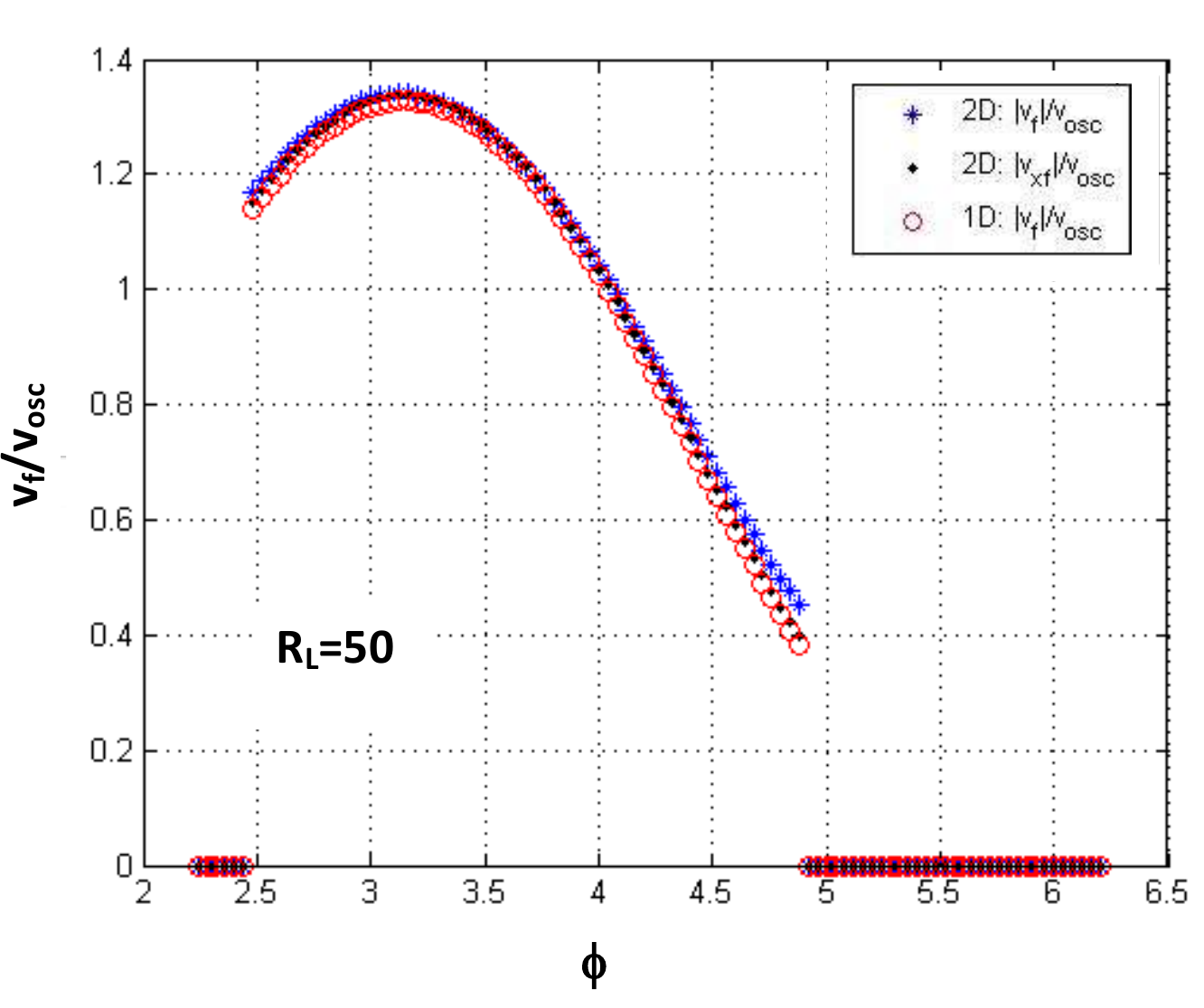}
\includegraphics[height=5.5cm,width=7.cm]{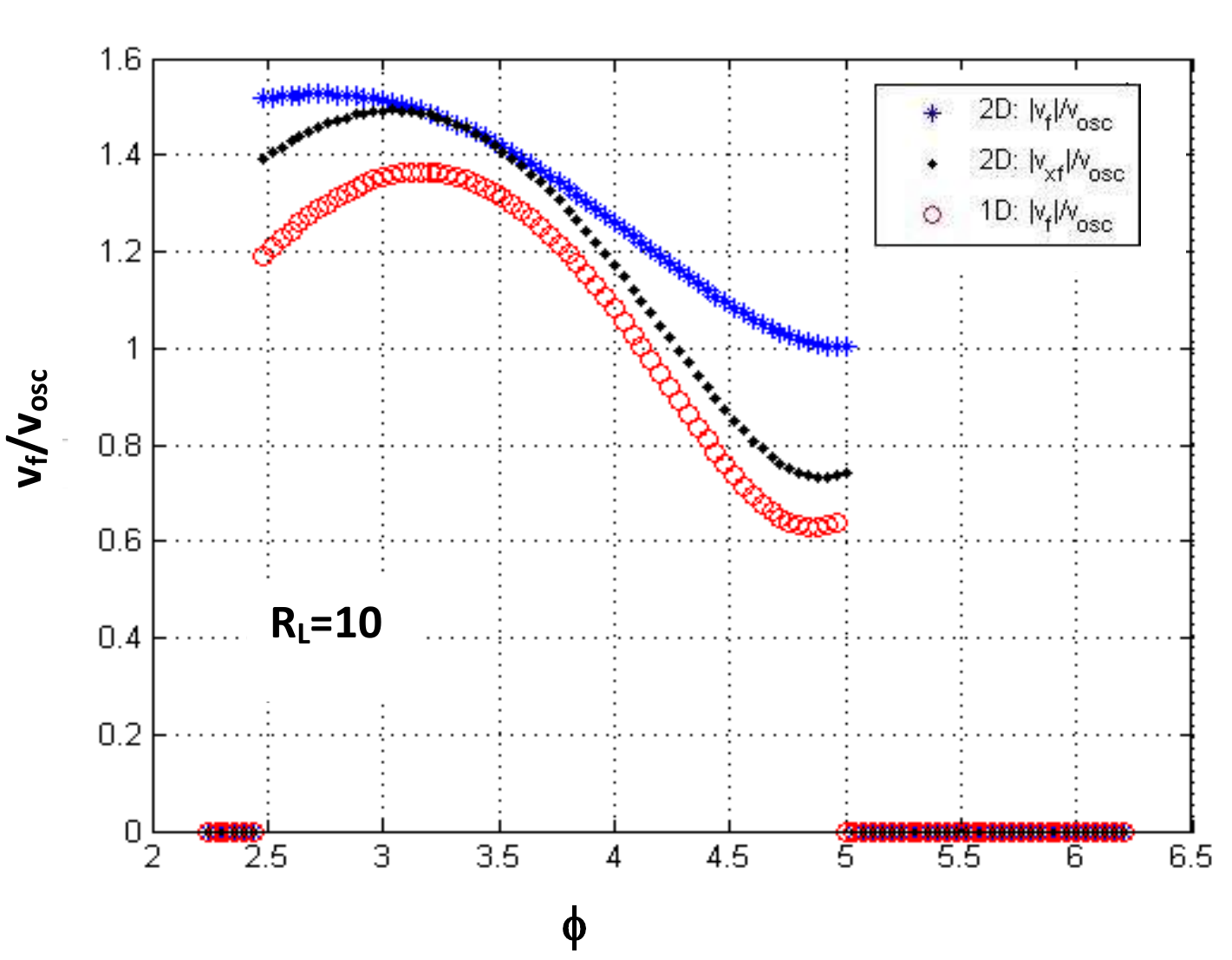}
\caption{
\label{fig:02} Comparison of the  final electron velocity  $v_f/v_{osc}$ from the $1D$ and $2D$ simulations as a function of the  phase $\phi$ at entry of  the surface plasma wave field for $R_L \sim 50$ ($\omega = 0.22 \omega_{pe}$) (upper part) and $R_L \sim 10$ ($\omega = 0.6 \omega_{pe}$)(lower part). {$\color{red}\circ$} $1D$ simulation, $\blacklozenge$ $x$ component of the momentum and {$\color{blue}\ast$} total momentum in the $2D$ simulation. Only the electrons in the vacuum side have been considered.} 
\end{figure}

In the electromagnetic regime where $\omega\ll\omega_{pe}$, taking into account the $E_y$ and $B_z$ components of the field is expected to have a minor effect on the particle motion for two reasons: i) The particle speed is always much smaller than the speed of light, c, and thus the magnetic-field contribution is small, and ii) $E_{y,v}/E_{x,v} \ll 1$. This can be observed in the upper part of fig.\ref{fig:02}  where for comparison we have reported  the  final electron velocity from the $1D$ and $2D$ simulations as a function of the  phase $\phi$ when the electron enters the surface plasma wave field for $R_L \sim 50$ ($\omega = 0.22 \omega_{pe}$). In this case  $E_{y,v}/E_{x,v}=0.2$. 
We can see that there is a very small difference between the $1D$ limit of the momentum 
and the $x$ component of the momentum in the $2D$ limit.  As in the one-dimensional case, we have a well-defined  shape for the velocity distribution as a function of the phase. The existence of a flat extremum around $\phi \sim \pi$ implies that many particles  will have the same value of the momentum despite their difference in phase. This results in a bunching of the electrons  in the direction perpendicular to the plasma surface in  momentum space, and a concomitant energy bunching\cite{ESW1}.

\begin{figure}
\includegraphics[height=3.8cm,width=4.27cm]{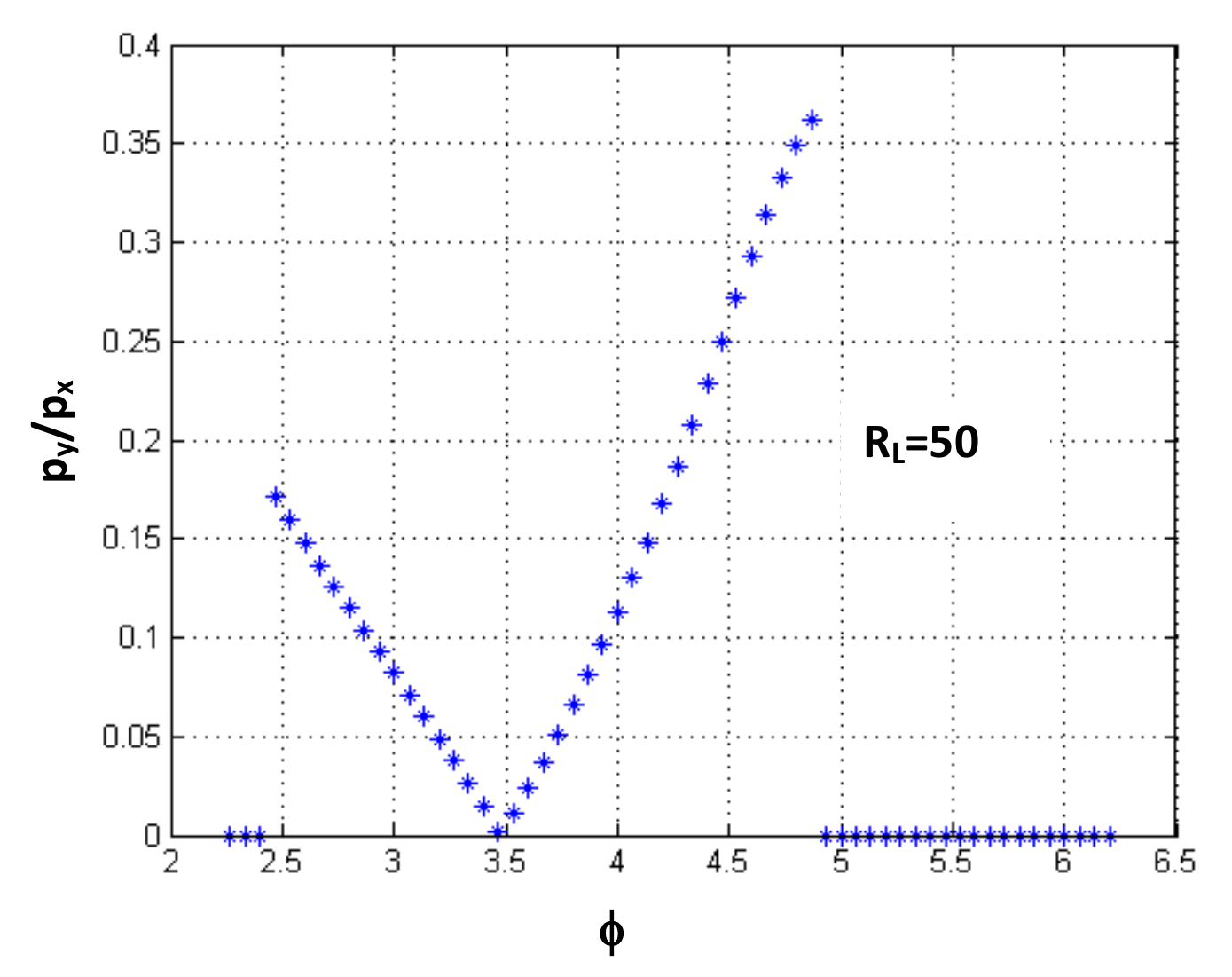}
\includegraphics[height=3.8cm,width=4.27cm]{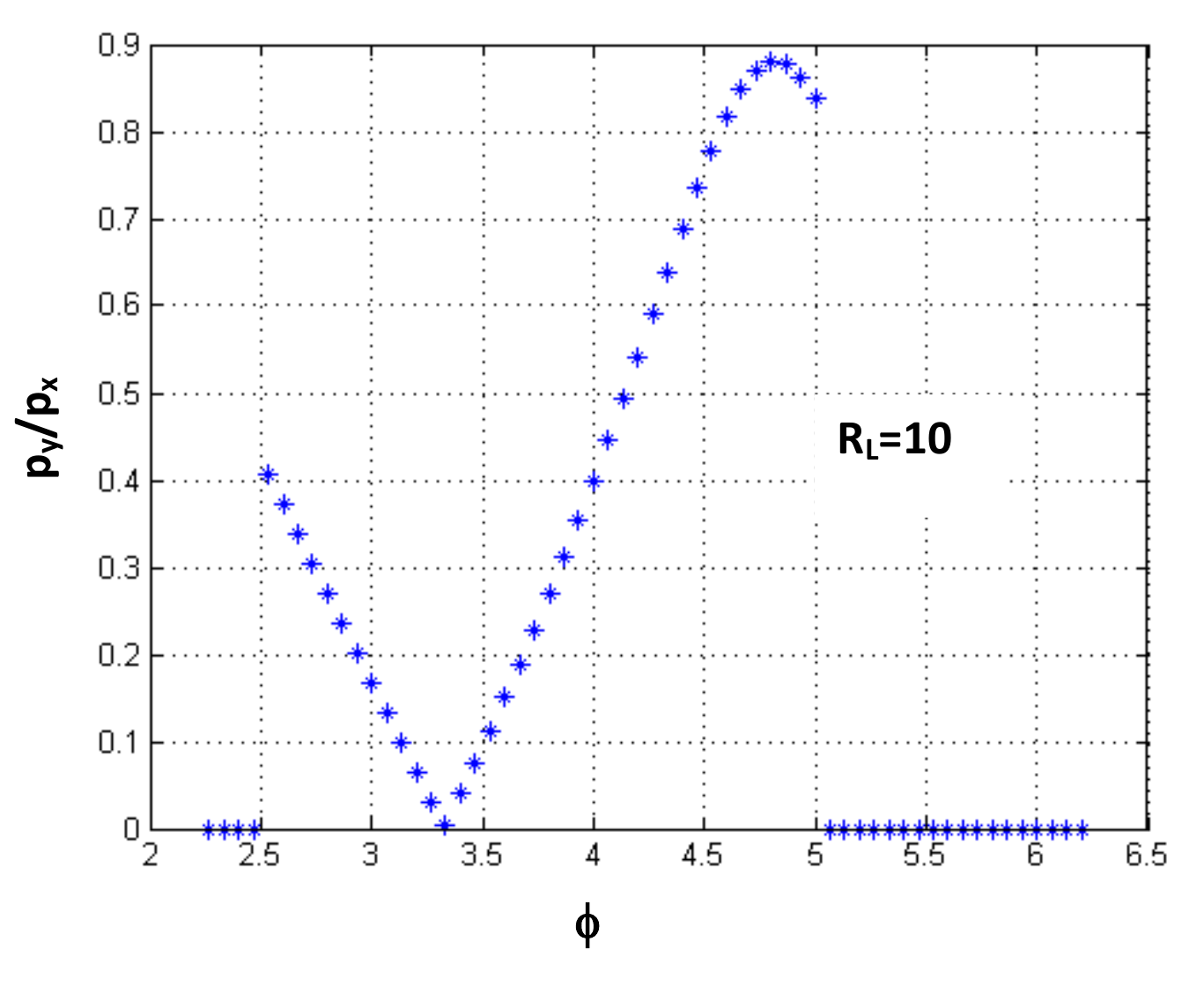}
\caption{
\label{fig:03} Evolution of the ratio $p_y/p_x$ of the  final electron momentum component for the $2D$ case as a function of the  phase $\phi$ at entry of the surface plasma wave field for $R_L \sim 50$ ($\omega = 0.22 \omega_{pe}$) (left) and $R_L \sim 10$ ($\omega = 0.6 \omega_{pe}$)(right). Only the electrons in the vacuum side have been considered. } 
\end{figure}

Increasing $\omega/\omega_{pe}$ we enter  the electrostatic regime where $2D$ effects may become important. For $R_L \sim 10$ ($\omega = 0.6 \omega_{pe}\sim \omega_{pe}/\sqrt{2}$) we have $E_{y,v}/E_{x,v}=0.75$. A significant difference between the velocity in the $1D$ case and the $x$ component of the velocity in the $2D$ case can now be observed as displayed in the lower part of fig.\ref{fig:02}. The analysis of the ratio $p_y/p_x$ of the  final electron momentum  for the $2D$ case as a function of the entry phase $\phi$ (see fig.\ref{fig:03} (right)) shows that  the weight of $p_y$ is especially important for $4<\phi<5$. The $1D$ description is then no longer valid  since $p_y$ becomes of the same order of magnitude as $p_x$. With respect to the estimate based on the ponderomotive potential Eq. (\ref{kabal}), the overall velocity acquired by the electron is enhanced by an amount that varies between $15\% $ and $60\%$ (depending on the phase).  
In the left part of fig.\ref{fig:03},   we have for comparison also plotted  the ratio  $p_y/p_x$  
in the electromagnetic limit ($\omega = 0.22 \omega_{pe}$, $R_L \sim 50$). It can be observed that in this limit $p_y$ is always significantly smaller  than $p_x$. 
We point out that in the case $R_L\sim 10$,  the final momentum acquired by the electron is higher than in the $1D$ case, even if the final $p_y$ is close to zero for phases between 3 and 3.5. This is a result of the history of the particle motion as at earlier times $p_y$ exhibits an oscillatory behavior that influences also the motion in the $x$ direction. 

To summarize the case of low surface plasma field intensity, the electron motion  
within the electromagnetic regime is essentially $1D$ in the  
direction perpendicular to the surface. If the SPW lifetime is higher than $L_{E,v}/v_{osc}$, the electrons with a favorable initial phase will gain more energy than can be accounted for on the basis of the ponderomotive contribution
only. In the electrostatic regime, the energy gain will be even higher ($\sim 25 \%$ increase) due to $2D$ effects. 

\section{Relativistic Regime}

We consider now very high field intensities such that $a_{sw}=eE_{sw}/mc\omega > 1$.
The electrons can then be accelerated to relativistic velocities.
 (Here $E_{sw}$ is the maximum value of the electric field in the $x$-direction in vacuum and we have taken it for  reference field). The electrons will then have a behavior in the field that varies, depending
 on the issue if  for $a_{sw}\gg1$ the field evanescence length $L_{E,v}$ becomes comparable or shorter than the characteristic high-frequency electron motion length $\Lambda=v_{osc}/\omega \sim c/\omega$. In this limit various other parameters become important such as the wave phase velocity, $v_\varphi$, and the relative value of the two components of the electric wave field at the surface, ${E_{y,v}}/{E_{x,v}}$. 

In this section we first present a parametric study (section IV.A) of the energy transfer to the electrons.  Only the transverse field of the evanescent surface wave is thereby considered. Within this $1D$ approach it is possible to make some analytical estimates which allows  identifying  the different interaction regimes  and the role of the evanescence length. The full $2D$ simulations follow in section IV.B, where the importance of the acceleration in the direction of the propagation of the surface wave will become evident.

\subsection{ Acceleration perpendicular to the surface: role of the evanescent length }

Due to the high values of the surface wave field and the very short  life time of the modes under consideration, we have here $v_{osc}\gg v_0$, where $v_0$ might be in the range of $\sim 100 eV-1 keV $.  We therefore do not expect our results to have any significant dependence on the initial value of the electron velocity $v_0$, which is therefore neglected in the following analytical estimate.

In the relativistic electrostatic limit where $\omega\sim\omega_{pe}/\sqrt{2}$, we can give an analytical estimate of the momentum acquired by the electrons as a function of the entry phase, since $R_L \ll 1$. In this case, an electron can acquire a large velocity $v \lesssim v_{osc}\sim c$ over a time span  less than a period, such that $\Lambda \gg L_{E,v}$. This means that the electron will leave the field and move towards the vacuum before 
the field will have had the time of going through an oscillation. 
The maximum energy gained by the electron under this hypothesis is given by $W_{K,f} = e E(\phi) L_{E,v}$, since we can assume that the electron sees a roughly constant value of the electric field $E(\phi)$ during the time it spends in the field of the surface plasma wave. Consequently, the value of the normalized electron momentum $p_f/m v_{osc}$ in this regime is given by (for $\pi < \phi < 2 \pi $):

\begin{equation}\label{kabal2}
{p_f \over m c} ~=~ \left[\left(1 + {p_{osc} \over m c}{ 2 \pi L_{E,v} \over \lambda_0} \cos (\phi + {\pi\over 2} ) \right)^2 -1 \right]^{1/2}.
\end{equation}

\begin{figure}[h!]
\includegraphics[height=5.5cm,width=7.cm]{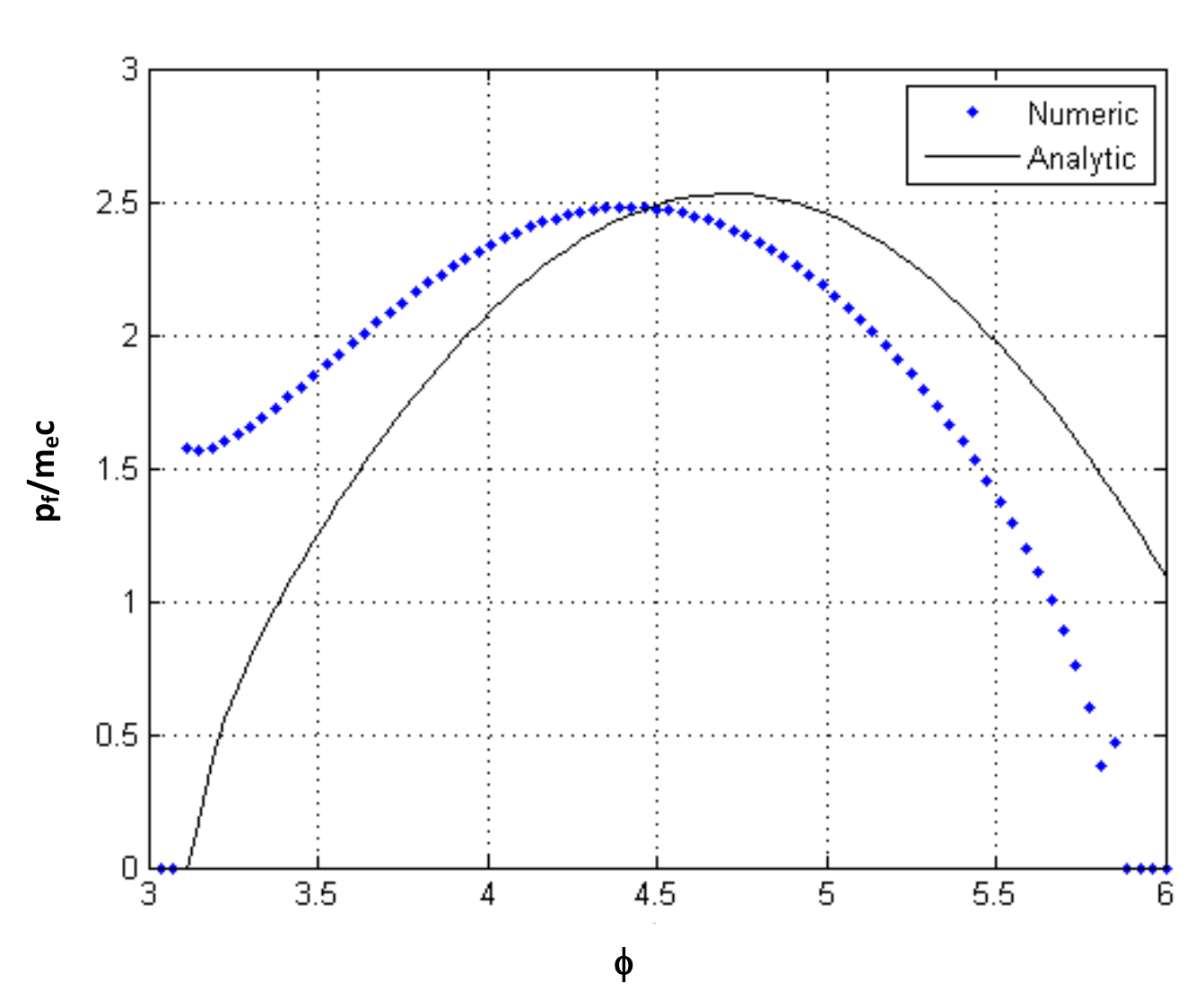}
\caption{
\label{fig:04}  Final electron momentum $p_f/(m c)$ as a function of the entry phase in the surface plasma wave field for $R_L=0.2$ ($\omega= 0.7 \omega_{pe}$)(the black ful line is the analytic prediction Eq.\ref{kabal2}). Only the electrons in the vacuum side have been considered.}
\end{figure}

The validity of this formula has been checked numerically for the case $R_L=0.2$ ($\omega /\omega_{pe} = 0.7$), for which the hypotheses leading to the equation above apply. In fig.\ref{fig:04} we plot $p_f/mc$ versus the entry phase for $R_L=0.2$ (
$\omega=0.7\omega_{pe}$, $L_E=0.032\mu m$, $a_{sw}=eE_{sw}/mc\omega=8.6$) and a surface plasma wave lifetime $\tau_{sw}$  of $3\tau_{0}$, where $\tau_{0}=2\pi/\omega$ is the period of the wave. 
Notice that the result presented in the figure is in fact basically independent of the SPW lifetime, since the electron spends less than a period in the SPW field. As we see, the analytical formula Eq.(\ref{kabal2}) can be considered as a fairly good estimate of $p_f$ versus $\phi$. In particular it provides with very good accuracy the maximum value of $p_f$ that can be obtained by an electron in this regime characterized by $R_L \ll 1$, and thus the {\it{lower limit}} for particle acceleration at a given wave amplitude compared to optimum 2D situation. In that sense it can be used as a reference case,  also owing to the fact that the energy transfer has a simple analytical formula.

When the scale of variation of the field is of the same order or larger than the distance explored by the electron in the field during one period, we have $R_L \sim \gg 1$. We denote in the following this situation as the relativistic electromagnetic regime. In this case, the kinetic energy acquired by an electron depends strongly on the phase of the field. As the possibilities of an analytical treatment are limited here, we present a numerical investigation of the final value of the momentum as a function of the entry phase. In the following we consider values of $R_L$ ranging from $0.9$ to $20$, and we solve  the relativistic equation of motion of electrons entering in the oscillatory evanescent field of the SPW, as defined in section III.

First of all, the assumption made that in this relativistic regime the details of the initial distribution function of the electrons do not affect the final range of momentum acquired by the electrons leaving the field, has been checked. This assumption is already well verified for $a_{sw}= 2.72$ and $T_e = 100 eV-1 keV$. Consequently, an uncertainty about the initial temperature of the electrons at the plasma surface will not significantly affect their final energy. In the following we take thus a constant value of $v_0$ corresponding to $100 eV$.
 
In this section we have taken electrons entering  the field during one period only, at the moment 
that the time envelope of the surface wave is at its peak (t=0). Under this hypothesis the only effect of the surface plasma wave lifetime  is reducing the amplitude of the field seen by the electrons during their motion towards the vacuum region, as 
was already observed in the 
non-relativistic case. As $v_{osc}$ is now close to the velocity of light, we have in most cases $ L_{E,v}/v_{osc} \tau_{sw} \ll 1$, such that we can neglect  the effect of the finite lifetime of the surface plasma wave in the discussion. In all the simulations reproduced here its value is taken equal to $\tau_{sw}=30\tau_{0}$.

\begin{figure}
\includegraphics[height=5.5cm,width=7.cm]{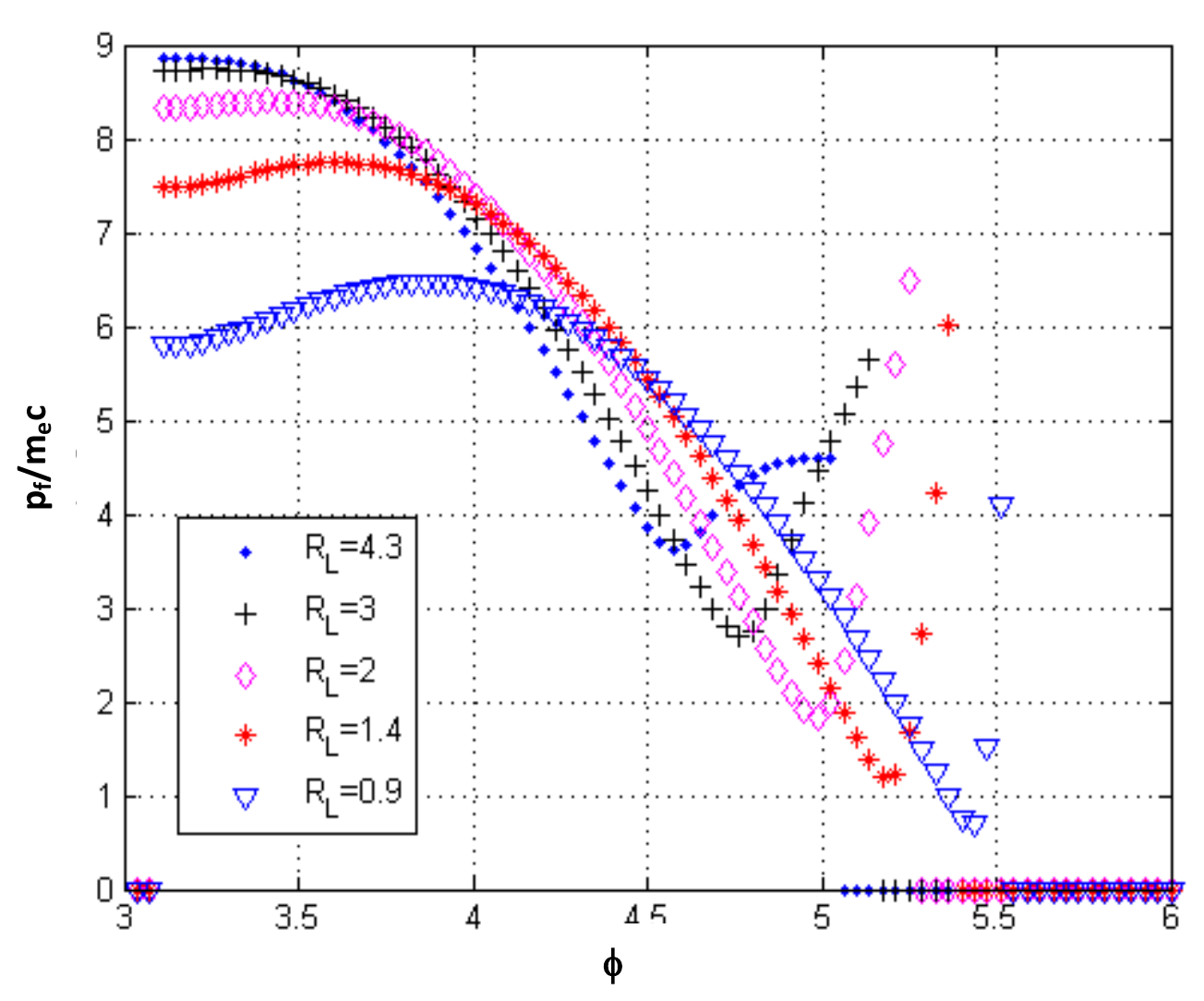}
\caption{\label{fig:05} Final electron momentum $p_f/(m c)$ as a function of the entry phase in the surface plasma wave field. Various values of $\omega/\omega_{pe}$ are considered. By varying this parameter we explore the values from $R_L=4.3$ ($\omega/\omega_{pe} =  0.22 $) to $R_L=0.9$ ($\omega/\omega_{pe} = 0.6 $), where the latter value tends towards the electrostatic regime. Only the electrons in the vacuum side have been considered.}
\end{figure}

\begin{figure}
\includegraphics[height=5.5cm,width=7.cm]{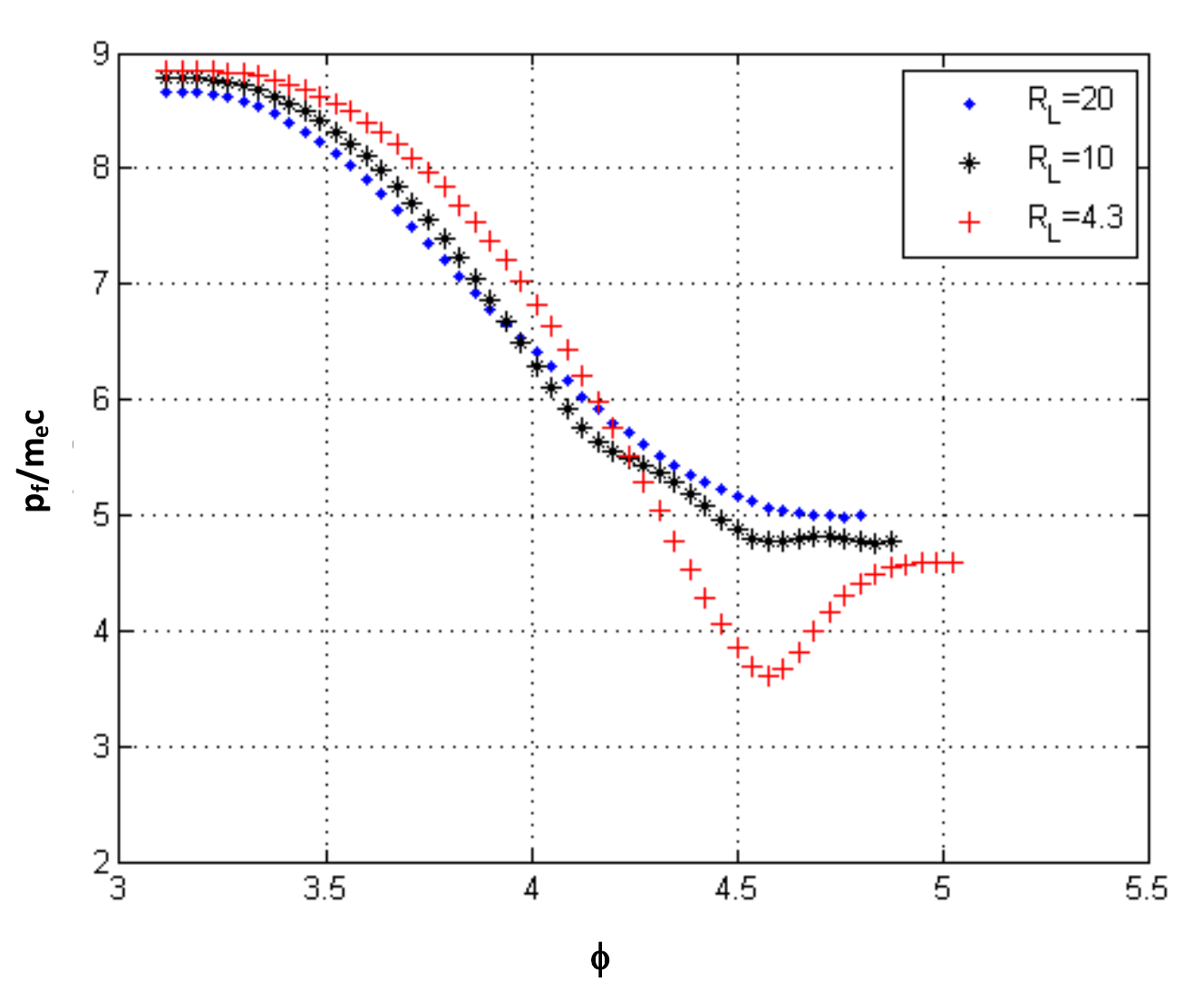}
\caption{
\label{fig:06}  Final electron momentum $p_f/(m c)$ as a function of the entry phase in the surface plasma wave field going from $R_L=4.3$ ($\omega/\omega_{pe} =  0.22 $) to $R_L=20$ ($\omega/\omega_{pe} =  0.05$). Only the electrons in the vacuum side have been considered.}
\end{figure}

We consider hereafter first the strong surface wave field limit with $a_{sw}= 8.6$ and values of $R_L$ ranging from $0.9$ to $4.3$ (fig.\ref{fig:05}). In a second stage we consider  values from $4.3$ to $20$ (fig.\ref{fig:06}) in order to fully explore the relativistic electromagnetic regime. Notice that  in the relativistic limit the quantity $R_L = L_{E,v} ~\omega/v_{osc}$, depends only on the ratio $\omega/\omega_{pe}$. In fact, since $v_{osc} \rightarrow c$, we have, using the surface plasma wave dispersion relation, eq. (\ref{disp}) $R_L \sim L_{E,v} ~\omega/c =\omega_{pe}/\omega \sqrt{1 - 2 \omega^2/\omega_{pe}^2}= \gamma_{\phi} / \sqrt{K} \sim \gamma_{\phi}$.

As it can be seen from fig.\ref{fig:05}, the maximum value of the momentum $p_f$, that can be acquired by an electron, increases steadily with $R_L$. The maximum value is obtained for $R_L=4.3$, for which we have $L_{E,v} \lesssim \Lambda \sim c/\omega$, and $p_{f_{max}} \gtrsim p_{osc}$. It can also be seen that an electron with an entry phase close to or larger than $3 \pi /2$ will acquire a larger momentum
 than those with neighboring values for the phase, resulting in a minimum, and then a local maximum. This can be related to the fact that such electrons make a turn after traveling in the field gradient, such that instead of being pushed back into the plasma they 
 start moving away towards the-low field region. When 
$R_L$ is close to one, this acts like a sort of resetting of the initial conditions, as illustrated in ref. \cite{ESW1}.

Let us now consider the results from fig.\ref{fig:06}, where we have plotted $p_f/(m c)$ versus the entry phase for larger values of $R_L$. It is clear from the figure that the maximum value of $p_f$ does not change much as $R_L$ increases. The maximum energy transferred to an electron diminishes slightly with respect to the case 
where $R_L >~ 1$, $L_{E,v} \sim c/\omega$, for which $p_f \gtrsim p_{osc}$, but as $R_L$ gets larger we still have $p_f \sim p_{osc}$. The ``hills'' that appear in the figure for $\phi$ close to $3\pi /2$ for the case $R_L=10$ are again due to particles that perform a turn in vacuum before leaving the high-field region, a single turn for particles in the first ``hill'' or two turns for the particles that form the second ``hill''. However we see that  there are no such ``hills'' if $R_L=20$. This is because now the scale of variation of the field is much larger than the path length covered by an electron over a single period, such that the particles either are pushed back into the plasma or perform many turns before leaving the high-field region. In this case the effect of ``resetting'' the initial conditions which is able to
produce a local increment of $p_f$ around $\phi = 3 \pi/2$, does no longer show up. 

To summarize: In the relativistic regime, the limit $R_L \gtrsim 1$, $L_{E,v} \sim c/\omega$  corresponds to a much {\it{more efficient  mechanism to convert the energy}} from the plasma wave to the electrons than in the opposite limit $R_L \ll 1$. We have not reported here the results for finer variations of $\omega/\omega_{pe}$, but as long as $L_{E,v} $ is close to $\sim c/\omega$, the maximum value of $p_f$ does not change significantly. Moreover, the condition $R_L \gtrsim 1$  is necessary to obtain an efficient acceleration of the particle  along the surface, as this condition permits the particle to stay for a long time close to the surface and to interact with the field parallel to the surface. This will be shown in the next section.

\subsection{2D simulations}

The quantity ${E_{y,v}}/{E_{x,v}}$, reflects the relative weights of the field components in 
the parallel $y$  and the perpendicular $x$ directions. From an inspection of its values reported in table I, we expect that the influence of the parallel component of the field on the electron motion will vary according to the regime  considered. In this section, we are thus going to investigate the full 2D motion of the particle. The simulations have been run taking the finite lifetime of the surface plasma wave equal to $\tau_{sw}=50\tau_{0}\sim 314\omega^{-1}$. 

\begin{figure}[h!]
\centering
\includegraphics[height=5.5cm,width=6.5cm]{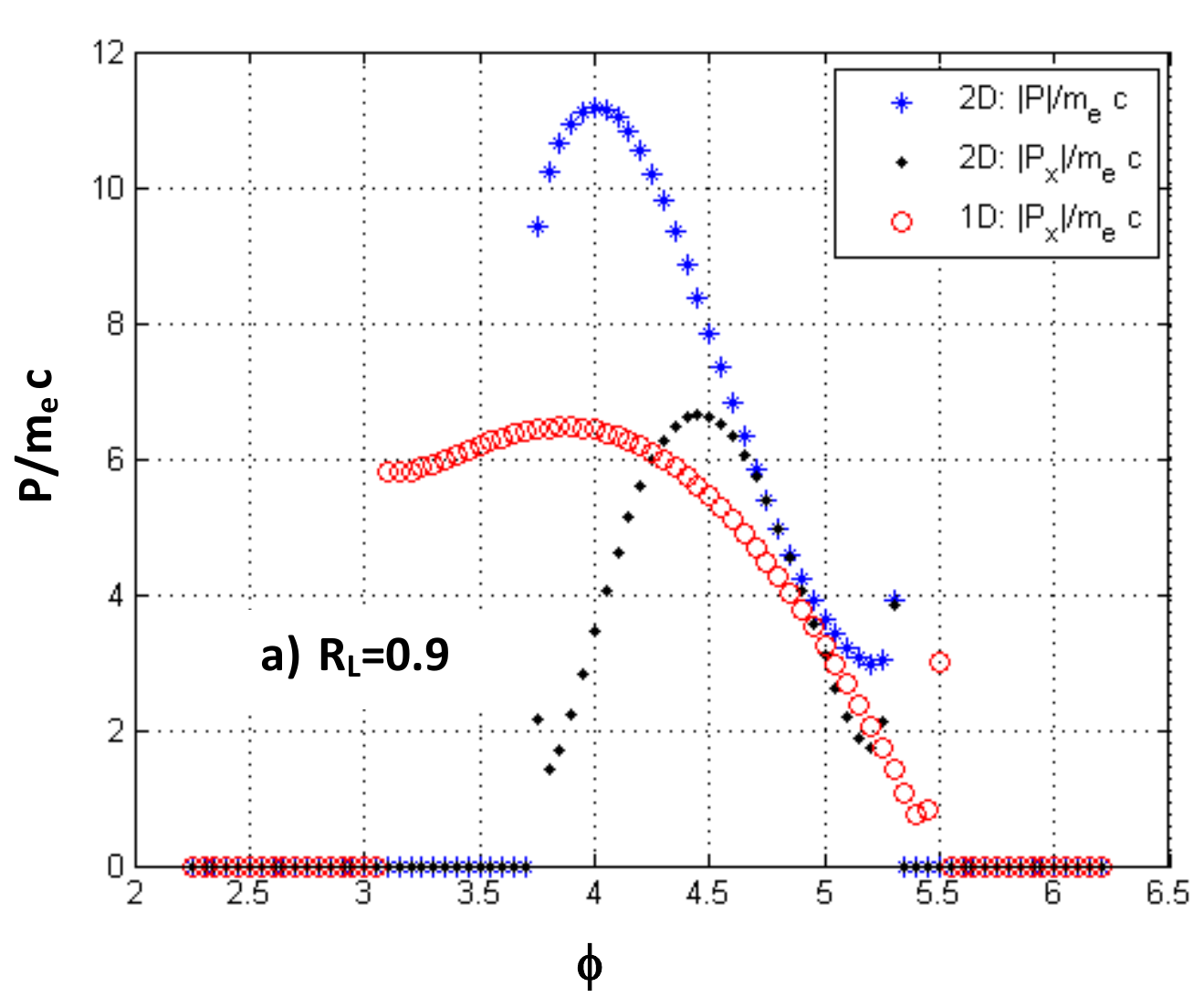} 
\includegraphics[height=5.5cm,width=6.5cm]{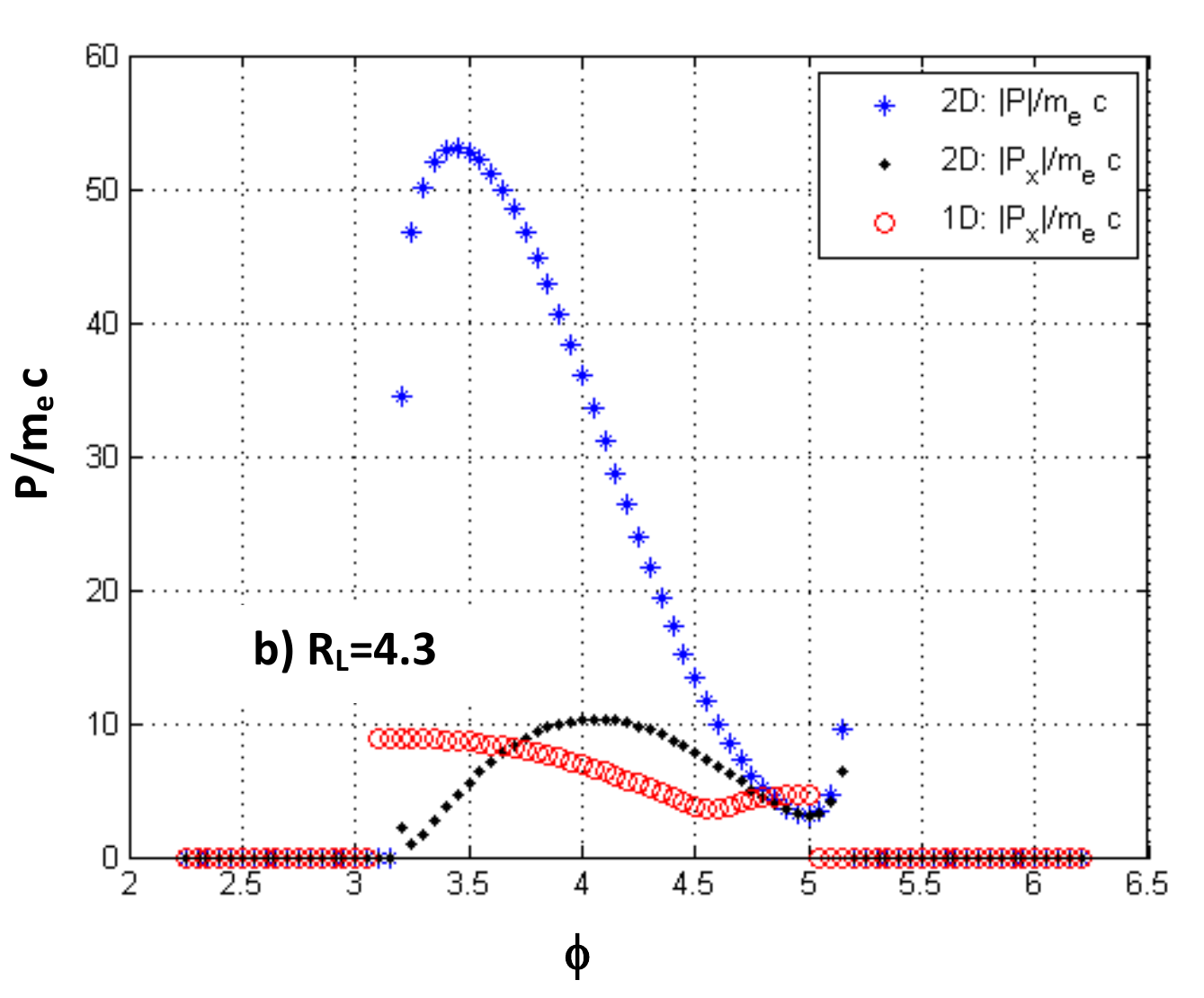} 
\includegraphics[height=5.5cm,width=6.5cm]{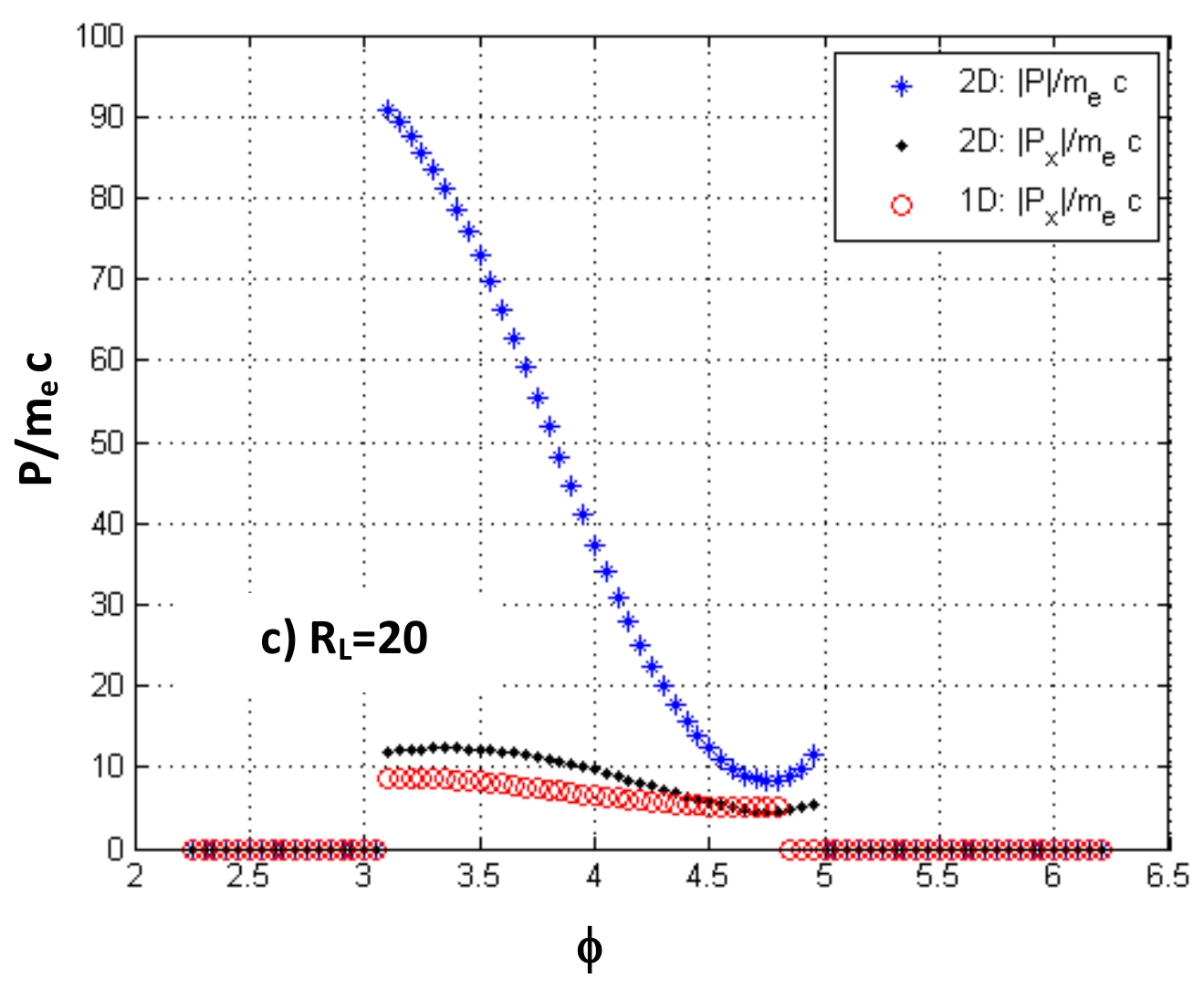}
\caption{
\label{fig:07} Comparison of the final electron  momentum for $a_{sw}=8.6$ from the $1D$ and $2D$ simulations as a function of the entry phase $\phi$ in the surface plasma wave field a) for $R_L=0.9$ ($\gamma_{\varphi}=1.33$, $\omega/\omega_{pe} =  0.6$) b) for $R_L=4.3$ ($\gamma_{\varphi}=4.54$, $\omega/\omega_{pe} = 0.22$) and c)for $R_L=20$ ($\gamma_{\varphi}=20$, $\omega/\omega_{pe} =  0.05$). {$\color{red}\circ$} $1D$ simulation, $\blacklozenge$ $x$ compoment of the momentum and {$\color{blue}\ast$} total momentum in the $2D$ simulation. Only the electrons in the vacuum side have been considered.}
\end{figure}

We first consider {\it{the relativistic electrostatic case}}, corresponding to $R_L \ll 1$ and $\gamma_{\varphi} \sim 1$. As found in the previous section, this is the case  for which the energy transfer mechanism is less efficient as the particle spends only a very short time in the perpendicular field. However, the two-dimensional effects lead to a final total momentum that is larger than in the case where only the transverse field is considered.  
This can be observed in fig.\ref{fig:07}a where we reproduce the total final electron momentum for the case $R_L = 0.9$ ($\omega /\omega_{pe} = 0.6 $ and $\gamma_\varphi=1.33$). This figure also displays the value of the  component of the final electron momentum along the perpendicular direction and the value for the momentum obtained from a 1D model. 
Like in the non-relativistic regime,
 the effect of the  electric field  parallel to the surface  is not negligible, especially for values of $\phi \sim 4$.
This can be seen from table I. The effect is of the same order magnitude as that of the field in the perpendicular direction.

 The maximum value of the momentum in the 2D case is roughly twice as large  as the value obtained in the previous section. An inspection of the ratio $p_y/p_x$ reveals that the motion parallel to the surface is no longer negligible
 in the case where the electron energy gain is maximal. Although  this motion contributes  to the enhancement of the maximal energy
 reached, the energy gain in this regime remains of the same order of magnitude as in 1D. 

\begin{figure}[h!]
\includegraphics[width=6.5cm]{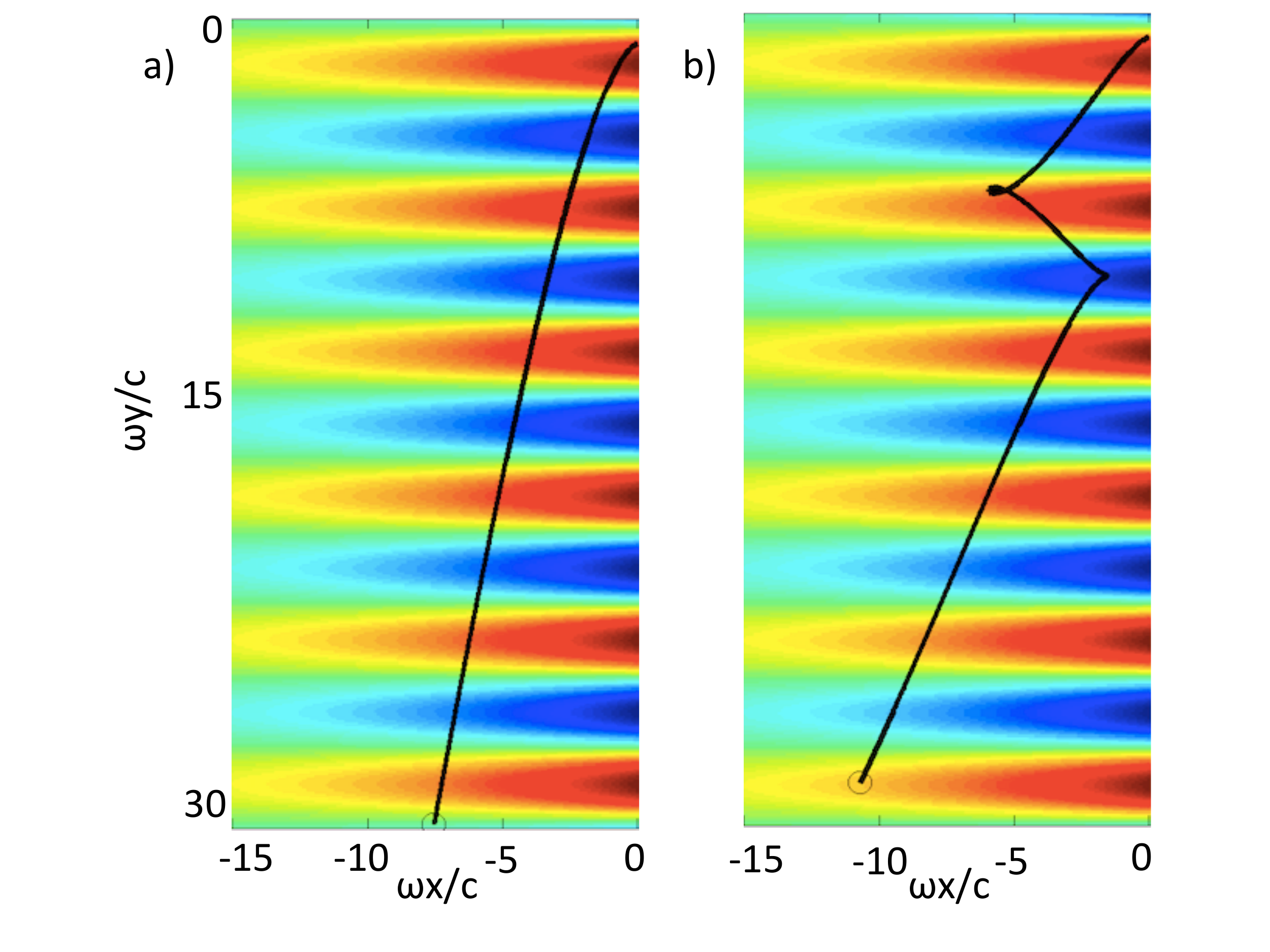}
\caption{
\label{fig:08} Trajectory of an electron in the surface plasma wave field from time $\omega t= 0.$ 
to $\omega t=40$ (before it leaves the accelerating region).  In a) the optimal entry phase $\phi=3.4$, 
that always sees an accelerating field (Multimedia view 
); in  b) the lowest energy entry phase $\phi=5.0$, that sees both an accelerating and decelerating field, acquiring 
mainly perpendicular energy  (Multimedia view 
). Parameters correspond to fig. \ref{fig:07}b.
The electric field $E_y$ at time $\omega t=40$ is superimposed to the trajectories for reference. }
\end{figure}

We next consider {\it{the relativistic electromagnetic regime}} wherein $R_L$ and $\gamma_{\varphi}$ 
increase from values $\gtrsim 1$ to values $\gg1$. As in the previous section (IV. A) we first take 
$R_L = 4.3$ ($\gamma_{\varphi}=4.54$ and $\omega /\omega_{pe} = 0.22 $). 
As  can be observed in fig.\ref{fig:07}b the  final value for the maximal electron momentum is significantly larger 
than the 1D value. From the fact that the $p_x$ component in the $2D$ case is very close to value of the momentum in the $1D$ 
case ($p_x\sim p_{osc}$) we can infer that most of the acceleration for the optimal entry phase  $\phi=3.4$ 
takes place in the parallel direction. This is corroborated by fig.\ref{fig:08}a : the trajectory of an electron for the 
phase $\phi=3.4$ from time $\omega t= 0.$ 
to $\omega t=40$ is reproduced superposed to the $E_y$ field of the surface plasma wave at the final time. 
In this case the electron  always sees an accelerating field (multimedia view). 
The trajectory of an electron in the same time interval for the lowest energy entry phase $\phi=5.$ is shown in fig.\ref{fig:08}b. 
 In this case the electron  sees both an accelerating and decelerating field, acquiring mainly perpendicular energy (multimedia view). 
A similar behavior is observed for the case $R_L = 20$ in fig.\ref{fig:07}c.  
However, here the maximum energy is even larger (mainly due 
to an increase of the parallel momentum as we still have $p_x\sim p_{osc}$). 

\begin{figure}[h!]
\includegraphics[height=6.cm,width=6.5cm]{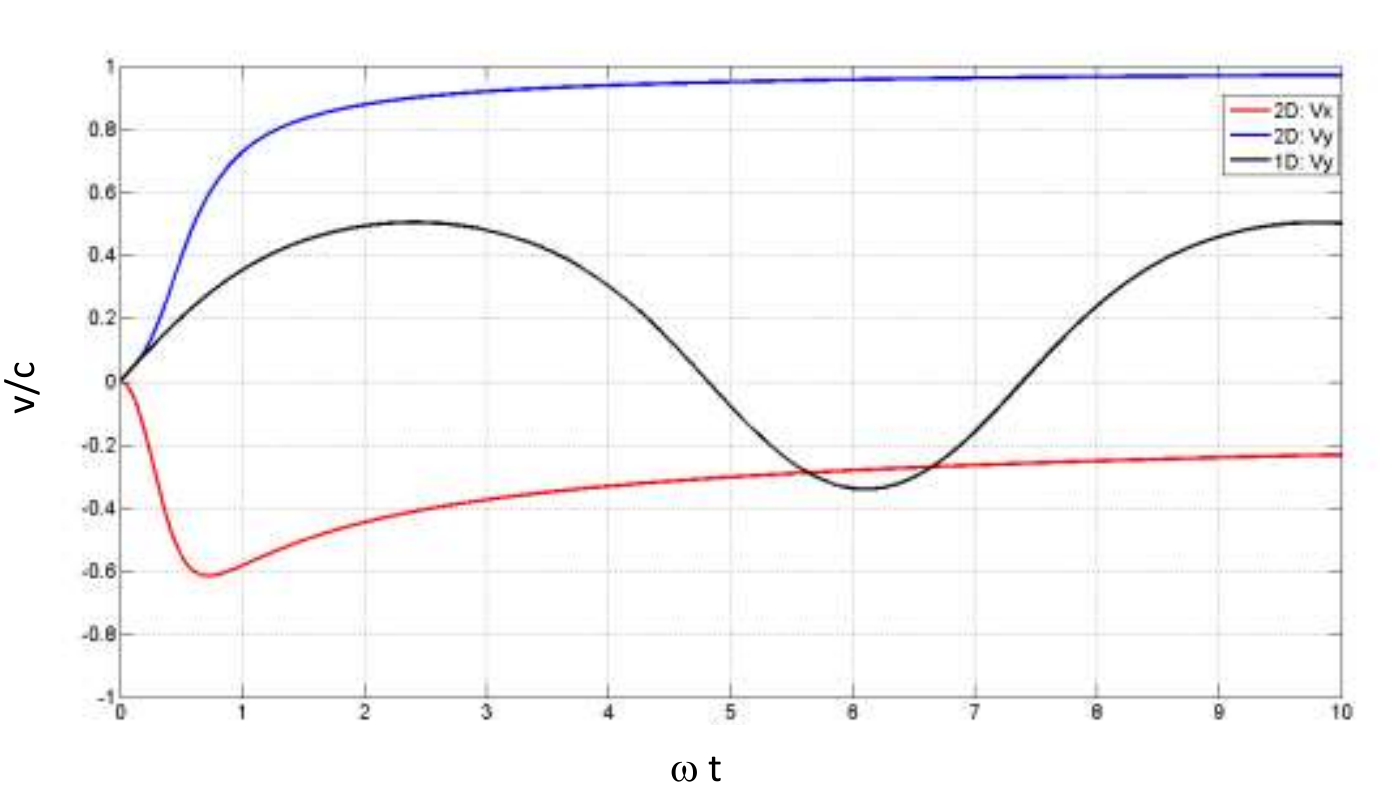}
\caption{
\label{fig:09} Evolution of the normalized electron velocity with time for the entry phase $\phi=3.14$ in the surface plasma wave field within the relativistic electromagnetic regime for $R_L=20$ ($\gamma_{\varphi}=20$ and $\omega/\omega_{pe} =  0.05$): $v_x$, $v_y$ ($2D$ simulation) and $v_y$ (setting $E_x=B_z=0$). Only the electrons in the vacuum side have been considered.}
\end{figure}

In order to accelerate an electron in the direction parallel to the surface by a wave that propagates along the surface it is necessary that the particle velocity be close to the wave phase velocity. This can be obtained by injecting  the electron right at the beginning with the appropriate velocity. 
Alternatively
an electron may be initially at rest, and then become self-injected  and phase-locked thanks to some other mechanisms. In our case, a significant contribution to the self-injection comes from the $v_x \times B$ force. This is shown in fig.\ref{fig:09} where we reproduce the $v_x$ and $v_y$ components of the electron velocity as a function of time for the phase, $\varphi\sim 3.1$, which gives the higher $p/m_ec$ value in fig.\ref{fig:07}c. For comparison we also plot the $v_y$ component for a case where $E_x$ and $B_z$ are set to zero, such that the force is absent. While in the latter case the electron oscillates in the field without gaining energy on average, we can see that  the particle gets an extra contribution in the y direction and that its velocity becomes close to $v_\varphi \sim c$ when the full 2D calculation is included.  

\begin{figure}[h!]
\includegraphics[height=6.5cm,width=8.5cm]{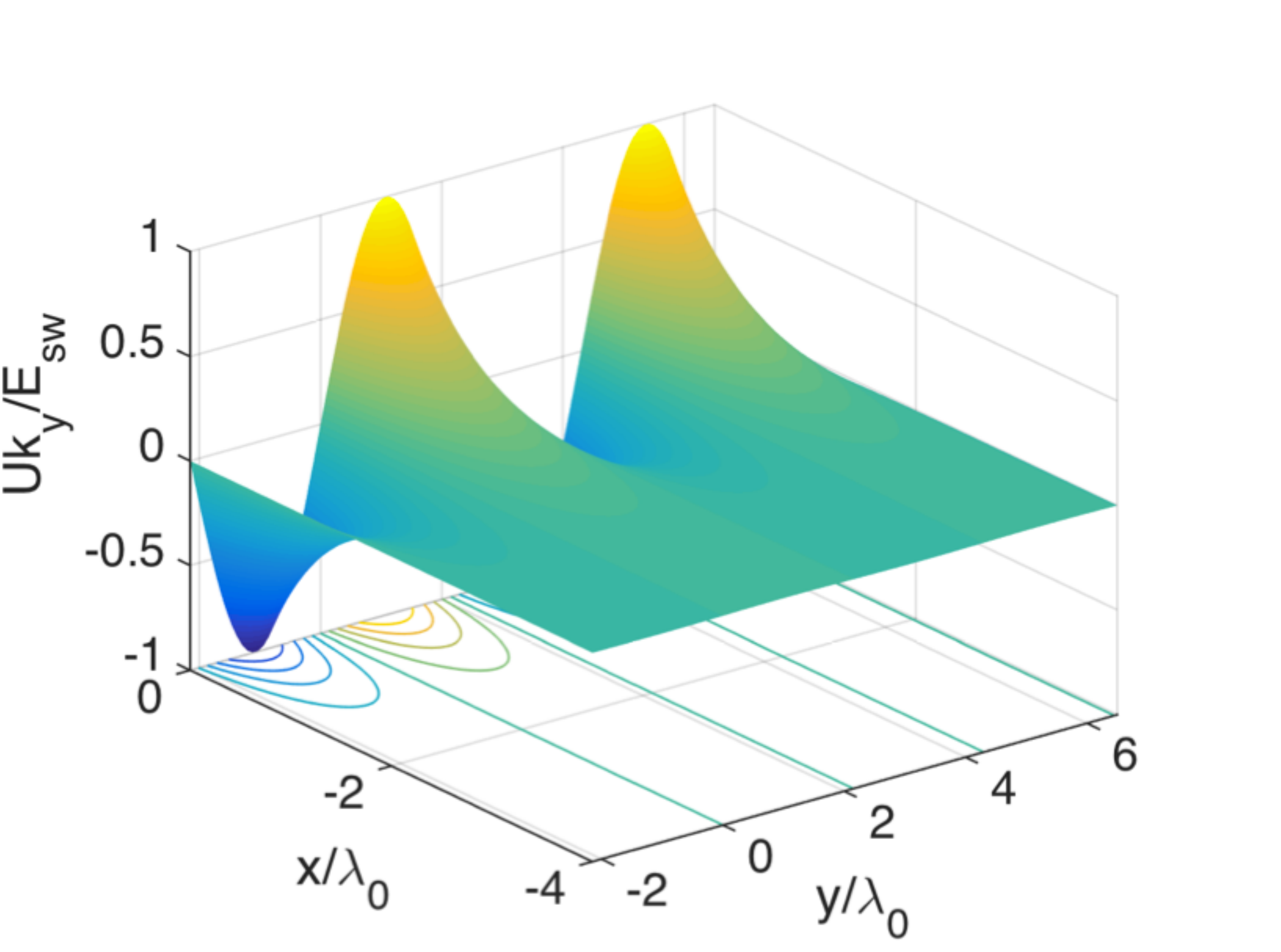} 
\caption{
\label{fig:10} Electrostatic potential [Eq. (11)] as a function of space for the case $R_L=4.3$ ($\gamma_{\varphi}=4.54$, $\omega/\omega_{pe} = 0.22$, $\phi=0$).}
\end{figure}

To further analyze the electron behavior in the SPW field, it is now useful to study the particle-field interaction in the reference frame of the wave phase velocity. To this extent, we perform a Lorentz transformation to the frame of the surface wave moving at $v_\varphi = \omega/k_y$. In this frame, we have:
\begin{equation}\label{ch-repere}
   \begin{cases}
       \omega t- k_y y =- k_y \tilde{y}/\gamma_{\varphi}, \\
       \tilde{t}=\gamma_{\varphi}(t- v_{\varphi} y / c^2),\\
       \tilde{\gamma}=\gamma \gamma_{\varphi}(1- v_{\varphi} v_y / c^2),\\
       \tilde{x}=x,\\
       \tilde{z}=z,\\
   \end{cases} 
\end{equation}
\noindent where the tilde refers to the quantities in the moving frame. 
When we neglect the lifetime of the surface plasma wave, the field expressions on the vacuum side Eqs.(\ref{champvide})  become:
\begin{equation}\label{ch_vid_b}
   \begin{cases}
       \tilde{E}_{x,v}= -E_{sw}\frac{h(\tilde{x})}{\gamma_{\varphi}} sin(-k_y \tilde{y}/\gamma_{\varphi}+\phi), \\
       \tilde{E}_{y,v}=  E_{sw} \frac{h(\tilde{x})}{k_y L_{E,v}} cos(-k_y \tilde{y}/\gamma_{\varphi}+\phi), \\
       \tilde{B}_{z,v}= 0,\\
   \end{cases} 
\end{equation}
with $h(\tilde{x})=\exp(\tilde{x}/L_{E,v})$ with $\tilde{x}<0$. \\
\\
It should be noted that after this Lorentz transformation we now are dealing with an electrostatic problem on the vacuum side in the moving frame, with $\tilde{E}_{x,v} \sim \tilde{E}_{y,v}$. The electrostatic potential, $U$, defined through $\overrightarrow{E}= -\overrightarrow{\nabla}U$, can be written as: 
\begin{equation}\label{pot}
U=-(E_{sw}/k_y) \exp(\tilde{x}/L_{E,v})sin(-k_y \tilde{y}/\gamma_{\varphi}+\phi),
\end{equation}
while the related potential energy, $W_p$, reads:
\begin{equation}\label{Epot1} 
 \frac{W_p}{mc^2}= a_{sw} \frac{v_{\varphi}}{c} \exp(\tilde{x}/L_{E,v}) sin(-k_y \tilde{y}/\gamma_{\varphi}+\phi).
\end{equation}
A plot of the electrostatic potential as a function of space in the surface wave reference frame is shown in fig. \ref{fig:10} for the case $R_L=4.3$ ($\gamma_{\varphi}=4.54$, $\omega/\omega_{pe} = 0.22$): clearly the initial conditions and in particular the initial phase seen by the electron in this potential strongly affect the possibility of gaining energy.

If we consider an electron at rest in the laboratory frame, in the boosted frame it will have a velocity $\tilde{v}_y=-v_\varphi$ and a normalized kinetic energy $W_K/(mc^2)=\tilde{\gamma}-1=\gamma_{\varphi}-1$. If the potential energy is comparable to the kinetic energy, $a_{sw} v_{\varphi}/c \sim \gamma_{\varphi}$, then for
an appropriate choice
of its initial phase, the electron can be efficiently slowed down in the $y$ direction in the boosted frame (and thus accelerated in the laboratory frame). Meanwhile it can also gain some perpendicular momentum by exploring the gradient of the potential in the perpendicular direction. The net final kinetic energy of the electron after leaving the potential has to be of the same order of magnitude of the potential itself:
\begin{equation}
\frac{W_{K,f}}{mc^2}= \sqrt{1+ \frac{\tilde{p}^2_{f,x}}{m^2c^2}+ \frac{\tilde{p}^2_{f,y}}{m^2c^2}}-1\sim \chi a_{sw} v_{\varphi}/c,
\end{equation}
 where $\chi<2$, while we note the final electron  momentum components as $\tilde{p}_{f,x}$ and $\tilde{p}_{f,y}$. It should be noted that the limiting value $\chi=2$ corresponds to the case where the electron has fully explored the potential in the $y$ direction. It  can never be reached in a two-dimensional situation due the $x$ component of the SPW field.

In the laboratory frame, the total energy of the electron is given by:
\begin{equation}
\frac{W_{TOT}}{mc^2}=\gamma_\varphi \left(\chi a_{sw} v_{\varphi}/c + 1 + \frac{v_\varphi \tilde{p}_{f,y}}{mc^2}\right), 
\end{equation}
with:
\begin{align}
p_x & = \tilde{p}_x, ~~~~~~~~~~~~~~~~~~~~~~~~~~~~~~~~~\\
p_y & = \gamma_\varphi \left( \tilde{p}_y + m v_\varphi(\chi a_{sw} v_{\varphi}/c + 1 )\right).
\end{align}

\noindent Hence in the limit $a_{sw}>1$ and  $v_\varphi \sim c $ we have:
\begin{equation}\label{Wtot}
\frac{W_{TOT}}{mc^2}=\gamma_\varphi \chi \left(a_{sw} + \frac{\tilde{p}_{f,y}}{mc}\right). 
\end{equation}

\noindent The final energy of the electron in the laboratory frame thus depends on the value and sign of $\tilde{p}_{f,y}$. 
The best situation would be obtained if $\tilde{p}_{f,y}>0$ and as large as possible, such that the electron  fully explores the potential in the $y$ direction, provided the potential barrier is high enough. In particular, in the limiting case where all the energy is in the parallel direction, we will have $\frac{\tilde{p}_{f,y}}{mc} \sim a_{sw}$ and the final energy will be equal to the well-known result for the energy gain in a sinusoidal wave \cite{mora} $4\gamma_\varphi a_{sw}$ (note that here we have $a_{sw}=eE_{sw}/mc\omega$, where $E_{sw}$ is the $x$ component of the field and $E_{y,v}\sim \gamma_\varphi E_{x,v}$).
However, as we already mentioned, this limiting situation cannot be reached due to the two-dimensional nature of the potential, which  will make the electron move away from the surface in the perpendicular direction. It is worth noting that the larger the phase velocity, the longer the time  the electron can spend  accelerating within the parallel field before  overcoming the wave. Consequently, the effect of the finite lifetime of the wave will be important 
in the higher phase velocity case. An upper limit for the time it takes to overcome the wave can be estimated as\cite{esarey} $\tau_d\omega=2\pi\gamma_\varphi^2$. Thus $\tau_{sw}$ should be comparable to $\tau_d$.

We can rewrite  Eq.(\ref{Wtot}) as:
\begin{equation}
\frac{W_{TOT}}{mc^2}=\chi^\prime\gamma_\varphi a_{sw},
\end{equation}
where depending on the value and sign of $\tilde{p}_{f,y}$, $\chi^\prime$ can be larger or smaller than $\chi$. As shown in fig.\ref{fig:07}b-c, the perpendicular momentum for the electrons with the highest energies is roughly given by $p_x/mc \sim a_{sw}$. Thus going back to the boosted frame and considering energy conservation we obtain $\chi \sim 1$. If $\tilde{p}_y>0$, we have $\chi^\prime > 1$ which corresponds to the case $\omega/\omega_{pe} = 0.22$. In this case the numerical value for the normalized final total momentum of the most energetic electrons is higher than $\gamma_\varphi a_{sw}$, as seen in table II. If instead $\tilde{p}_y<0$, we have $\chi^\prime < 1$ which corresponds to the case $\omega/\omega_{pe} = 0.05$. 
In this case, the numerical value for the normalized final total momentum of the most energetic electrons is smaller than $\gamma_\varphi a_{sw}$, as can be seen in table II.  

\begin{table}
\caption{\label{runs} Parameters for the different simulations discussed in the paper, $p_f/mc$ is the normalised final total momentum obtained for the most energetic electron in the simulation (see fig.\ref{fig:07}) .\\} 
\begin{tabular}{|c|c|c|c|c|}
\hline
$\omega/\omega_{pe}$ & $a_{sw}$ & $\gamma_{\varphi} a_{sw}$ & $p_f/m_ec$\\
\hline
0.22 &  8.6  & 40 &  52\\
     &  20.  & 91 &  130\\
\hline
0.05 &  8.6  & 172 &  91\\
     &  20. & 400 & 250 \\
\hline
\end{tabular} 
\end{table}

\begin{figure}[h!]
\centering
\includegraphics[height=5.5cm,width=6.5cm]{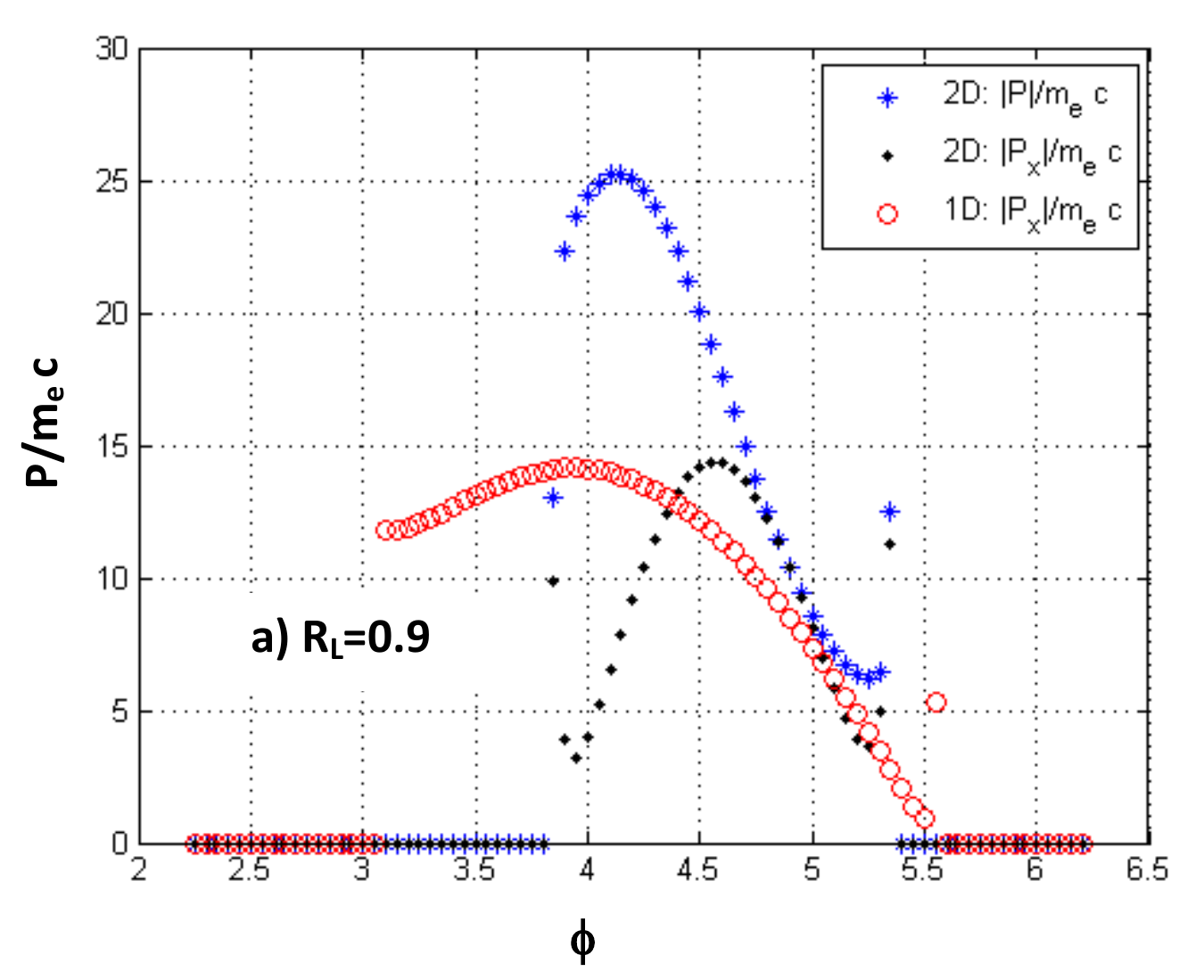} 
\includegraphics[height=5.5cm,width=6.5cm]{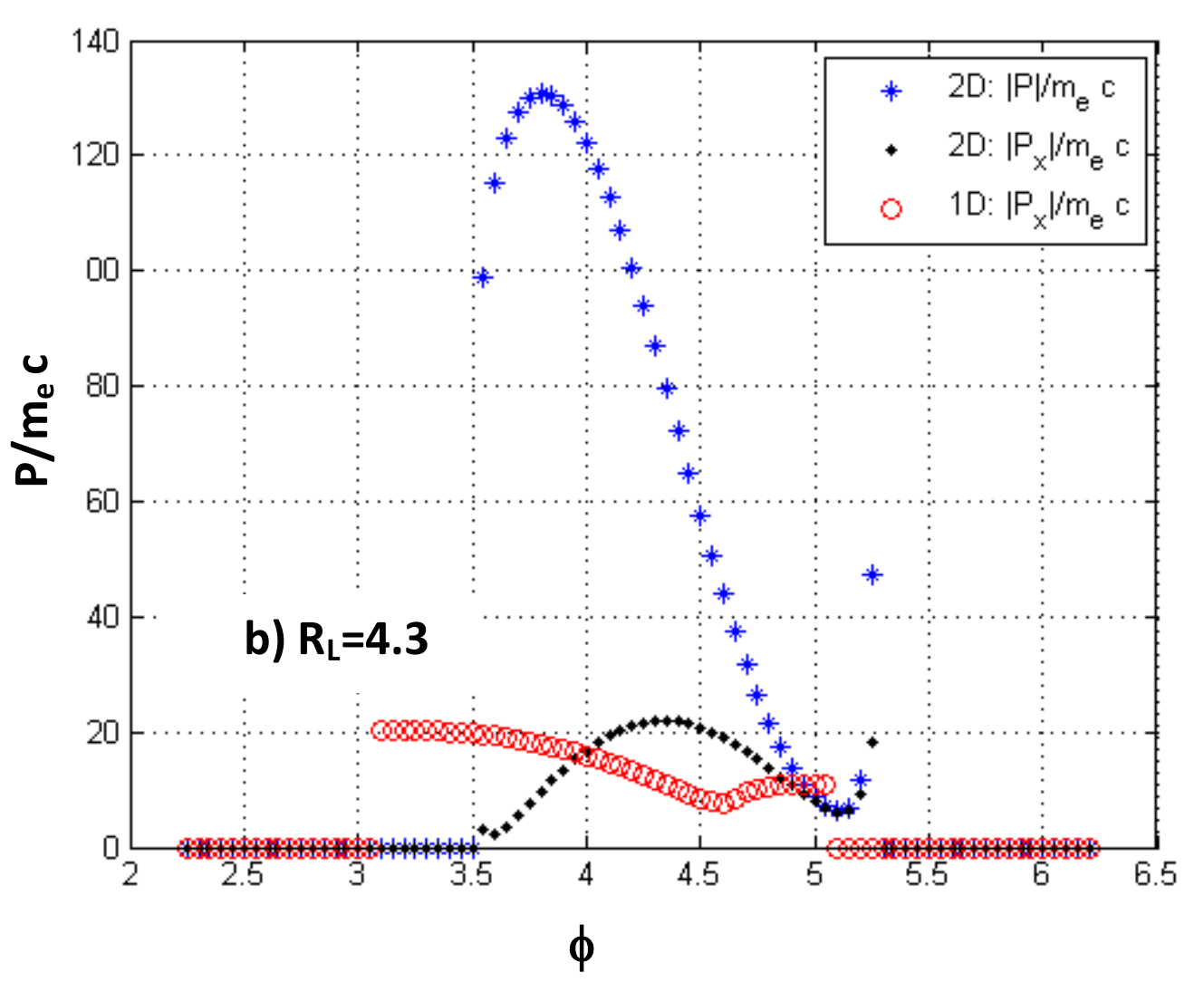} 
\includegraphics[height=5.5cm,width=6.5cm]{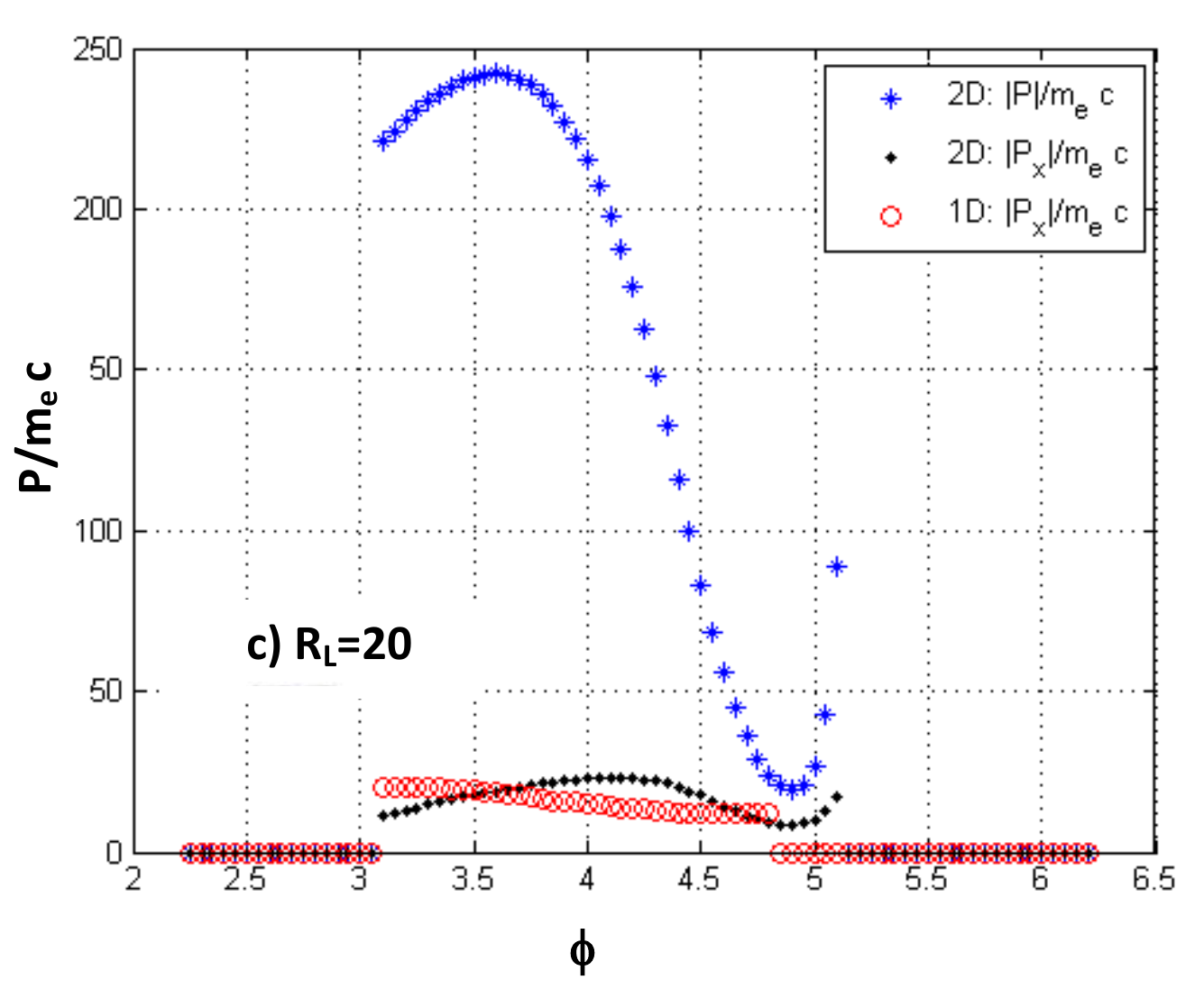}
\caption{
\label{fig:11} Comparison of the  final electron momentum in the case $a_{sw}=20$ from the $1D$ and $2D$ simulations as a function of the entry phase $\phi$ in the surface plasma wave field a) for $R_L=0.9$ ($\gamma_{\varphi}=1.33$, $\omega/\omega_{pe} =  0.6$) b) for $R_L=4.3$ ($\gamma_{\varphi}=4.54$, $\omega/\omega_{pe} =  0.22$) and c) for $R_L=20$ ($\gamma_{\varphi}=20.$, $\omega/\omega_{pe} =  0.05$). {$\color{red}\circ$} $1D$ simulation, $\blacklozenge$ $x$ compoment of the momentum and {$\color{blue}\ast$} total momentum in the $2D$ simulation. Only the electrons in the vacuum side have been considered.}
\end{figure}

In order to verify the scaling of the energy gain with $a_{sw}$ we now consider the simulation for $a_{sw}=20$. The final  electron momentum for this case and  the three situations previously studied is reported in fig.\ref{fig:11}. As seen in  table II, the scaling is confirmed: the normalized final total momentum for the most energetic electrons at $\omega/\omega_{pe} =  0.22$ gives a value for $\chi^\prime\ $ $=$ $1.4$ (instead of $1.3$ in the case $a_{sw}=8.6$). 
As previously pointed out, the finite lifetime of the surface plasma wave also affects the value of the normalized final total electron  momentum, as $\tau_d\omega$ is much larger than $\tau_{sw}\omega$ for  $\omega/\omega_{pe} =  0.05$.
As a matter of fact, increasing the lifetime by a factor $2$ for this case enhances the normalized final total momentum up to $p_f/mc=313$ ($\chi^\prime\ = 0.78$). 

Finally, a comparison for the electrostatic case is also given in fig.\ref{fig:11} and as expected the final momentum is here much less than in the other cases.

\section{Conclusion}

In this paper we present different regimes for electron acceleration  within the evanescent field of a surface plasma wave, which were obtained by means of  simulations of the motion of test particles. 

For low surface plasma field intensities (characterized by $a_{sw}\ll1$), we show that
 in the electromagnetic regime (characterized by $\omega \ll\omega_{pe}$)  the electron motion is 
essentially $1D$. 
The electron travels  into the vacuum in the direction perpendicular to the surface. This is due to the fact that the vacuum evanescence length $L_{E,v}$ of the SPW field is large compared to the amplitude of the high-frequency motion of the electron in the wave, $v_{osc}/\omega$. If the SPW lifetime is larger than $L_{E,v}/v_{osc}$, the electrons with a favorable initial phase will gain more energy than provided by the ponderomotive contribution. In the electrostatic regime where $\omega\sim\omega_{pe}\sqrt{2}$, the $E_y$ SPW field components induce positive $2D$ effects, which amplify the electron energy gain.

In the regime where relativistic effects are no longer negligible (for $a_{sw}>1$), the relative weight of the perpendicular versus parallel SPW field components may limit the interaction of the electron with the full two-dimensional field of the surface wave. Nevertheless, we find optimal conditions for acceleration in which the electron can still gain a large amount of energy, mainly by acceleration parallel to the surface.  
This is achieved when the electrons acquire a parallel velocity close to the phase velocity, $v_\varphi$, of the wave propagating along the surface. This is observed in the so-called relativistic electromagnetic regime where $\gamma_\varphi= 1/\sqrt{1 - (v_\varphi/c)^2}$ is large and $\omega \ll\omega_{pe}$. Reaching this optimal regime also depends on the lifetime of the surface plasma wave. 

In this regime an electron initially at rest is self-injected and phase-locks on the vacuum side. A significant contribution to the self-injection is provided by the $y$ component of the $\vec v \times \vec B$  force $(v_x B)$ resulting in favorable $2D$ effects in contrast to a purely electrostatic wave. This way  the electron gains a substantial amount of parallel energy. In the boosted frame on the vacuum side the $B$ field becomes  equal to zero while $E_{y,v}$ is invariant. However, even if in this frame the particle is not initially in phase with the wave and $\gamma_\varphi > a_{sw} v_\varphi/c$, the electron can still gain a large amount of energy. This happens while it  is subjected to the parallel field during the accelerating part where it is rolling away from the surface before the field changes its sign and becomes decelerating. This confirms the favorable effect of two-dimensional dynamics.
On the plasma side (this case has not been presented here) this mechanism is less efficient because the perpendicular component of the SPW electric field is smaller than in vacuum. Also its spatial extension is smaller. Moreover, on the plasma side the problem does not become electrostatic in the boosted frame.

We point out that an optimum energy gain could  also be  reached for external injected electrons with an high initial velocity and an appropriate angle of injection.

Finally, we want to outline that we have examined only electrons that are accelerated and are spending their dynamical history on the vacuum side. Depending on the phase, some electrons can also spend their dynamical history inside the plasma. 
Since the electric field perpendicular to the surface is discontinuous and changes sign we found that some electrons can oscillate back and forth from the surface to the plasma. After some oscillations they may stay on one side or another. However, we decided not to study such motion in the present work, since the numerical study cannot be precise in the presence of a discontinuity. Moreover, a more realistic situation including plasma temperature effects will easily smooth such a discontinuity, without otherwise significantly changing the surface wave solutions \cite{ale}. Other important effects that are missing in this simple model, but have been observed - e.g. in PIC simulations \cite{ale} -  are 
the presence of a charge space field at the plasma surface left by the electrons ejected from the surface, or the presence of a quasi-static magnetic field \cite{alePRE,naseri:13}. Considering these effects is beyond the scope of this paper and will be the subject of further studies.

\vspace{0.5cm}
One of us (C.R.) would like to acknowledge fruitful discussions with F. Amiranoff, A. Macchi, P. Mora and D. Pesme.

\end{document}